\documentclass[10pt,journal,cspaper,compsoc]{IEEEtran}
\ifCLASSOPTIONcompsoc
  \usepackage[nocompress]{cite}
\else
  \usepackage{cite}
\fi
\ifCLASSINFOpdf
\else
  \usepackage[dvips]{graphicx}
\fi
\usepackage[cmex10]{amsmath}

\usepackage{array}

\usepackage{mdwmath}
\usepackage{mdwtab}

\usepackage[tight,footnotesize]{subfigure}
\usepackage{url}

\usepackage{multirow}
\usepackage{booktabs}
\usepackage{amsfonts}
\usepackage{amssymb}
\usepackage{lscape}
\usepackage[ruled]{algorithm2e}
\usepackage[]{algorithmic}
\usepackage{rotating}
\usepackage{multicol}

\def\argmin{\mathop{\rm argmin}}

\newcommand{\eg}{{e.g.\  }}
\newcommand{\ie}{\emph{i.e.\ }}

\usepackage{framed, color}
\definecolor{shadecolor}{rgb}{1,0.5,0}

\newcommand{\Space}[1]{\ensuremath{\mathcal{#1}}}

\newcommand{\real}{\mathbb{R}}

\newcommand{\rcubed}{\real^{3}}
\newcommand{\rthree}{\rcubed}
\newcommand{\rnth}[1]{\real^{n}}

\newcommand{\s}{\ensuremath{\mathbb{S}}}
\newcommand{\stwo}{\ensuremath{\mathbb{S}^2}}
\newcommand{\sphere}{\stwo}
\newcommand{\sn}[1]{\s^{n}}
\newcommand{\domain}{D}

\newcommand{\surfaces}{\Space{F}}

\newcommand{\preshapes}{\mathcal{C}}
\newcommand{\preshapesf}{\preshapes_{f}}
\newcommand{\preshapesq}{\preshapes_{q}}
\newcommand{\srnfs}{\preshapesq}

\newcommand{\shapes}{\Space{S}}
\newcommand{\shapesf}{\shapes_{f}}
\newcommand{\shapesq}{\shapes_{q}}

\newcommand{\ltwo}{\mathbb{L}^{2}}

\newcommand{\diffeos}{\Gamma}
\newcommand{\diffeo}{\gamma}
\newcommand{\rotations}{SO(3)}
\newcommand{\rotation}{O}

\newcommand{\srnf}{q}

\newcommand{\meansurf}{\bar{f}}

\newcommand{\unitnorm}{\ensuremath{\hat{n}}}

\newcommand{\geodf}{{\alpha_{f}}}

\newcommand{\norm}[1]{\lVert #1 \rVert}

\newcommand{\inner}[2]{\langle #1,#2 \rangle}
\newcommand{\innerd}[2]{\langle \langle #1,#2 \rangle \rangle}

\newcommand{\tr}{\text{tr}}

\newcommand{\srnfmap}{Q}

\newcommand{\deriv}[2]{\dfrac{d #1}{d #2}}

\def\argmin{\mathop{\rm argmin}}

\def\disp{\textbf{d}}
\def\ffform{g} 
\def\shapeop{S}			

\def\normal{\text{n}}				
\def\realplus{\real^{+}}				

\def\trace{\text{tr}} 				
\def\det{\text{det}} 				

\def\energy  {E} 
\def\bendterm {\energy_{\text{b}}  } 		
\def\stretchterm  { \energy_{\text{s}} } 		
\def\cauchygreen {G}						

\def\path{F}

\def\qmap{Q}

\def\Hess{\text{Hess}}
\def \meancurvature{H}

\def\sphereone{\mathbb{S}}

\def\landmarkcount{l}
\def\nshapes{n}
\def\cov{K}

\def\no{no}
\def\yes{yes}

\graphicspath{{./}{figures/}{figures/cylinders/}{figures/classification/}{figures/medical/}}

\DeclareGraphicsExtensions{.pdf}

\begin{document}
\newcommand{\ra}[1]{\renewcommand{\arraystretch}{#1}}

%
\title{A Survey on Non-rigid 3D Shape Analysis}
%
%
%
%

\author{Hamid Laga~\IEEEmembership{}
\IEEEcompsocitemizethanks{\IEEEcompsocthanksitem Hamid Laga  is with the School of Engineering and IT, Murdoch University, Australia and with the Phenomics and Bioinformatics Research Center, University of South Australia.
E-mail: H.Laga@murdoch.edu.au.}
\thanks{Manuscript received April 19, 2005; revised December 27, 2012.}}

%
%

\markboth{A Survey on Non-rigid 3D Shape Analysis}%
{Laga \MakeLowercase{\textit{et al.}}: A Survey on Non-rigid 3D Shape Analysis}
%



\IEEEcompsoctitleabstractindextext{
\begin{abstract}
Shape is an important physical property of natural and manmade 3D objects that characterizes their external appearances. Understanding differences between shapes and modeling the variability within and across  shape classes, hereinafter referred to as \emph{shape analysis},   are fundamental problems to many applications,  ranging from  computer vision and computer graphics to biology and medicine.  This chapter provides an overview of some of the recent techniques that studied the shape of 3D objects that undergo non-rigid deformations including bending and stretching. Recent surveys that covered some aspects such classification, retrieval, recognition, and rigid or nonrigid registration,  focused on methods that use shape descriptors.   Descriptors, however, provide abstract representations that do not enable the exploration of shape variability. In this chapter, we focus on recent techniques that  treated the shape of 3D objects as points in some high dimensional space where  paths  describe deformations.  Equipping the space with a suitable  metric enables the quantification of the range of deformations of a given shape, which in turn  enables (1) comparing and classifying 3D objects based on their shape, (2) computing smooth deformations, \ie geodesics, between pairs of objects, and (3) modeling and exploring continuous shape variability in a collection of  3D models.  This article surveys and classifies recent developments in this field, outlines fundamental issues, discusses their potential applications in computer vision and graphics,  and highlights opportunities for future research. Our primary goal is to bridge the gap between various techniques that have been often independently proposed by different communities including mathematics and statistics, computer vision and graphics, and medical image analysis.


\end{abstract}

}

\maketitle

\IEEEdisplaynotcompsoctitleabstractindextext

%
\IEEEpeerreviewmaketitle

\section{Introduction}
\label{sec:introduction}

\bstctlcite{IEEEexample:BSTcontrol}
\IEEEPARstart{S}{hape} is an important physical property of natural and manmade objects that characterizes their external appearances. Understanding differences between shapes and modeling the variability within and across  shape classes  are fundamental problems and building blocks to many applications,  ranging from  computer vision and computer graphics to biology and medicine.  
In medicine, studying shapes of 3D anatomical structures and modeling their growth patterns are of  particular interest since many diseases can be linked to alterations of these shapes~\cite{Grenander:1998:CAE,Fishbaugh:2012:ALS,joshi2016surface}.   In computer graphics, 3D  modeling  using low level primitives (\eg  polygons)  is a tedious and time-consuming process. With the growing availability of 3D models  via online repositories,  recent research efforts have shifted their focus towards  data-driven techniques for generating  3D shape variations from existing ones.  Data-driven techniques are also gaining momentum in the field of 3D reconstruction and modeling from images, range scans and noisy point clouds~\cite{nan2012search}, which are ill-posed problems. By leveraging shared information across multiple objects, data-driven techniques aim at discovering patterns, which can be  used  as  
  geometric, structural,  and semantic priors for driving the 3D reconstruction and modeling processes.  

%


When dealing with numbers, statistics is the science for studying  and modeling variabilities. Geometric shapes are fundamentally different. Research in this area, driven by  the computer vision, medical imaging, computer graphics and statistics communities, seeks to develop mathematical tools for  manipulating shapes in the same way we manipulate numbers.  This includes:
\begin{itemize}
    	\item \textbf{Comparing the shapes of two objects,  }  \ie saying whether two objects have similar shape or not, and more importantly, quantifying and localizing  the similarities and differences.  
	\item \textbf{Computing summary statistics and prototypes,} such as  the mean shape, modes of variations and  higher order statistics, of a collection of 3D models. 
	\item \textbf{Mathematical modeling of shape variations,  }  using, for example, probability distributions as generative models. 
		These models form priors for random sampling and for  statistical inferences, and 
	\item \textbf{Exploring  the shape variations} for synthesizing  valid   shapes and performing  interpolations,   extrapolations, statistical inferences, and regressions.
\end{itemize}
Implementing these ideas, for 3D models,  requires solving several challenges, each one has been the subject of important research and contributions. 
The first one is about the representation  of shape.  When selecting any representation, it is important to  ensure that two different shapes cannot have the same representation and that a given representation can always be associated to a shape.

Second, almost any shape analysis task requires some measure of dissimilarity, hereinafter referred to as \emph{a metric},  for comparing shapes. The challenge when designing a suitable metric comes from the fact that 3D objects have arbitrary pose, scale, and parameterization. They also bend, stretch, and change their topology and structure.  Some of these variations, \eg  translation,  pose, scale, and parameterization, are shape-preserving, while others  affect the shape. Thus, the metric  should  be invariant to transformations that are shape-preserving  (translation, scale, rotation and re-parameterization) and should quantify  those that change the shape (bending, stretching, and structural changes).

The third difficulty in  the analysis of 3D shapes  is registration, \ie matching  points across objects.  Any shape metric requires a registration component to help decide which point on one object is compared to which point on the other.   This  is often a difficult problem to solve, especially across objects with large  deformations and pose variability. Thus, it is not surprising that a large body of literature is specifically dedicated to this problem, see for example~\cite{vankaick11correspsurvey} for a detailed survey of this topic.


Finally, there are several computation and numerical challenges that one needs to address. For instance, several shape representations are non-linear. This results in shape spaces and metrics that are non-Euclidean. As a consequence, standard techniques, such as Principal Component Analysis (PCA), for statistical analysis cannot be directly applied. Moreover, analytical solutions, \eg  for correspondence and geodesics, often  do not exist and one has to rely on optimizing non-convex potential functions. This makes the problem even challenging and one often has to compromise between accuracy and computational efficiency.


The purpose of this chapter is to provide a comprehensive survey of the state-of-the-art techniques in elastic 3D shape analysis. By elastic we refer to 3D shapes that undergo  elastic deformations, \ie those that bend and stretch. This already forms a large class of natural 3D objects  with a significant large body of literature and techniques dedicated to the topic.  We structure the survey into four main sections.  Section~\ref{sec:background} provides   a general formulation of the problem and some background materials on which the subsequent topics are built upon.  
Section~\ref{sec:shape_spaces} focuses on the concepts of shape spaces and metrics and their relation to physical deformations.
Section~\ref{sec:registration_geodeiscs} treats the elastic correspondence / registration problem and its closely related problem of computing geodesics between pairs of shapes. Section~\ref{sec:statistical_analysis}  discusses how these building blocks are used for computing mean shapes, studying the modes of variations and characterizing shape collections with statistical models. Section~\ref{sec:examples_applications} discusses a few concrete examples where these concepts are used. Finally, we conclude in Section~\ref{sec:summary} by summarizing the main lessons learned in this survey and discussing future directions for research.

%

\section{General formulation}
\label{sec:background}

We provide in this section a general formulation of the elastic shape analysis problem. We first overview the basic representations of 3D objects, which are essential for the subsequent sections of this chapter. Then, we formulate the elastic shape analysis problem and give the taxonomy that we will use for classifying state-of-the-art techniques.

\subsection{Representations}
\label{sec:representations}

Commonly, shape is interpreted as the boundary of a 3D object. Mathematically, it can be defined as a function $f: D \to \real^d$, which maps every   point of the domain $D$ to  a point in $\real^d$. Here $d$ is equal to $1$ or $3$ depending on the choice of the representation. 

A common choice of the function $f$, and thus the domain $D$, is the signed distance function, $f: D \subset \real^3 \to \real$, whose values at a point $s \in D$ is the signed distance of that point to the nearest point on the boundary of the object being represented. Positive (respectively negative) values of $f$ correspond to regions in space that are outside (respectively inside) the object.  Subsequently, points $s$ such that $f(s) = 0$ form the boundary of the object. This representation is commonly known as \emph{level-set} representation since for any  $\alpha \in \real$, the points $s$ such that $f(s) = \alpha$ form a surface called \emph{level-set}.   This  representation is implicit since one has to solve the equation $f(s) = 0$ in order to identify the points that belong to the object of interest.

In contrast to implicit representations, one can represent the boundaries of an object using an explicit function $f: D \to \real^3$, which assigns to every point $s \in D$ a three-dimensional point $f(s) = (x(s), y(s), z(s))$. The set of points $f(s)$ then forms the 3D object, which is often referred to as an embedding of the canonical domain $D$. Although this seems to be the most natural representation, the choice of the domain $D$ remains an open challenge since there is no single domain $D$ that is suitable for every type of objects. For instance, open surfaces such as 3D human faces can be easily embedded on a planar domain $D \subset \real^2$. $D$ can be, for example,  a unit disk. Similarly, closed surfaces of genus-0 can be embedded onto a unit sphere $\stwo$. 

But how is it done in practice? A 3D model is, in general,  produced either by scanning a real 3D object, using the various types of 3D sensors that are currently available, or by using a 3D modeling software. In any case, a 3D model is represented as an 
unorganized point cloud or a uniformly-sampled set of points. To facilitate rendering and visualization processes, the point clouds are converted into polygonal meshes since many rendering algorithms as well as graphics processing units are optimized for processing  planar, particularly triangular, polygons. As such, the boundary $f$ of a 3D object is represented as a piecewise linear surface  in the form of a  triangulated mesh $M$, \ie a set of vertices $P = \{p_1, \cdots, p_n\}$ and triangles $\mathcal{T} = \{T_1, \cdots, T_m\}$ such that the triangles  intersect only at common vertices or edges~\cite{floater2005surface}.  

Finally, the parameterization process takes a triangulated mesh $M$, finds a suitable continuous domain $\domain$, referred to as \emph{the parameterization domain},  and then estimates the underlying continuous surface $f: \domain \to \rthree$. If $M$ is a disk-like surface, then $\domain$ can be chosen as a simply-connected region  of the plane, \eg  the unit disk.  When dealing with closed genus-0 surfaces then the unit sphere $\stwo$ is the natural parameterization domain. We refer the reader to the paper of Floater et al.~\cite{floater2005surface} for a detailed survey of different parameterization techniques.

In what follows, we assume, unless explicitly specified,  that $\domain = \stwo$ and thus $s = (u, v)$ is composed of the longitude and latitude coordinates.

\subsection{Invariance requirements}
\label{sec:shape-preserving-transformations}

An important challenge, when designing a framework for shape analysis,  is to discard effects or variables  that do not affect the shape of an object and  to account only for those that affect shape.  
Kendall~\cite{kendall:1977DS} described shape as a property of an object once its rotation, translation,  scale, are parameterization are removed from the representation.
In the field of shape analysis, the transformations associated with translation, rotation, and uniform scaling are termed \emph{shape-preserving}. They are nuisance variables that should be discarded. Translation and scaling are probably the easiest ones to deal with.  For instance, let $\surfaces$ be the space of all surfaces $f: \domain \to \real^3$.  One can discard translation by first translating $f$ so that its center of mass will be located at the origin: 
$
	f(s) \rightarrow f(s) - \frac{\int_{\domain}a(s)f(s)}{\int_{\domain}a(s)}.
$
Here $a(s)$ is the local surface area at $s$. It is defined as the norm of the normal vector to the surface at $s$. In the case of spherically parameterized surfaces, i.e. $\domain = \stwo$ then $s=(u, v)$ and $n(s)= \frac{\partial{f}}{\partial{u}} \times   \frac{\partial{f}}{\partial{v}}$. The scale component can be also discarded by normalizing the surface $f$ in such a way that its overall surface area is one:
$
	f(s) \rightarrow  \frac{f(s)}{  \sqrt{  \int_D a(s)   }   }. 
$
The space  $\preshapesf$ of all origin-centered and unit-surface-area surfaces is called \emph{pre-shape space}.

In addition to scale and translation, 3D objects undergo two other shape-preserving transformations, which are rotation and re-parameterization. Rotations, denoted by $O$, are elements of $SO(3)$, which are  $3\times 3$ matrices. They transform each point $f(s) \in \real^3$ into $Of(s)$. Here, we denote by  $Of$ the rotated surface. 

\begin{figure}[t]
    \centering   

		 \includegraphics[width=.45\textwidth]{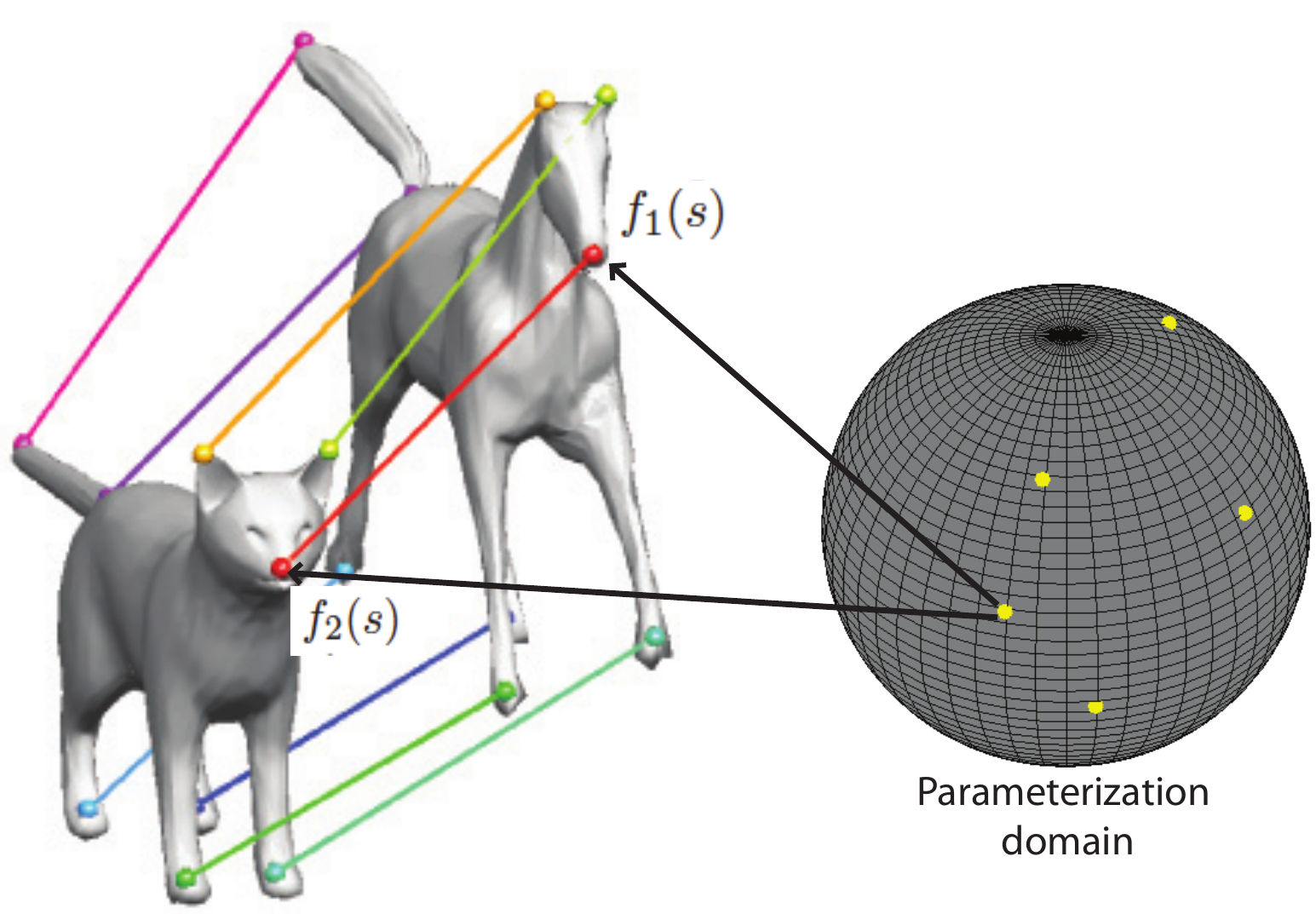}	 
    \caption{\label{fig:correspondence_example} An example of correct correspondences between two 3D objects that bend and stretch. In general, however, the analysis process should find the optimal reparamterization of $f_2$ to put it in full correspondence with $f_1$. } 
\end{figure}

Re-parameterization is a diffeomorphism $\gamma: \domain \to \domain$, which transforms a surface $f$ into $f\circ\gamma$. Let $\Gamma$ denote the space of all such diffeomorphisms. Diffeomorphisms are shape preserving transformations; that is, the surfaces  $f$ and $f\circ\gamma$ have the same shape, albeit having different parameterizations. Note that re-parameterization is important in 3D shape analysis because it provides registration. Consider two surfaces $f_1$ and $f_2$ where $f_1$ is, for example, the surface of a cat  while $f_2$ is the surface of a horse, eventually in a different pose, see Figure~\ref{fig:correspondence_example}. Let $s  \in \domain$ such that $f_1(s)$ corresponds to the nose tip of $f_1$. If  $f_1$ and $f_2$ are arbitrarily parameterized,  which is often the case since they  have been parameterized independently from each other, $f_2(s)$ may refer to any other location on $f_2$, the ear tip for example. Thus $f_1$ and $f_2$ are not in correct correspondence.  Putting $f_1$ and $f_2$ in correspondence is equivalent to reparameterizing $f_2$ with a diffeomorphism $\diffeo \in \diffeos$ such that for every $s \in \domain$, $f_1(s)$ and $f_2(\diffeo(s)) = (f_2 \circ \diffeo)(s)$ point to the same feature, \ie nose tip to nose tip, etc..

The simplest way of removing the rotation variability is to make sure that all shapes being analyzed are  aligned, in a consistent fashion, to a common coordinate system. One way to do this is to compute the principal axis of the objects being analyzed, using Principal Component Analysis (PCA), and align these axis to the axis of the coordinate frame. This approach, however, is very sensitive to outliers and is sub-optimal especially in the presence of large elastic deformations.

Alternatively, invariances to rotations and re-parameterizations can be dealt with  algebraically. The idea, in the case of rotations for example, is that a surface $f$ and any other surface obtained by rotating $f$ have the same shape. That is, $\forall \rotation \in SO(3)$, $f$ and $\rotation f$ are equivalent. Similarly, a surface $f$ and any of its possible reparameterizations $f \circ \diffeo$ are equivalent.   Thus the set $[f] = \{ \rotation f \circ \diffeo, \rotation \in SO(3), \diffeo \in \diffeos \}$ forms an equivalence class under the action of the rotation and reparameterization groups.  The set $\shapesf = \preshapesf / (SO(3)\times \diffeos) $ of all equivalence classes is called \emph{shape space}.

\subsection{Problem statement and taxonomy} 
\label{sec:elastic_shape_analysis_problem}
We are given a collection  of origin-centered and scale-normalized 3D objects $\{f_i \in \preshapesf, i=1, \cdots, n\}$. For the generality of formulation, we assume that each object $f$ is represented as an embedding: $f: \domain \to \real^3$, which assigns to every point $s \in \domain$ a 3D point $f(s) \in \real^3$.  When talking about shape analysis,  we are interested in (1)  quantifying similarities and differences in their shape, and (2) characterizing shape variability in the collection, which in turn can be done by computing summary statistics such as mean shape and modes of variation, and fitting probability distributions to the population in the same way as it is normally done with numbers.

Intuitively, the mean shape  can be defined as the object whose shape is as close (similar) as possible to all the shapes of the other objects in the collection. This is known as the \emph{Karcher} mean. Mathematically,  we seek to find a shape $\meansurf$ and a set of optimal rotations $\rotation^*_i$ and diffeomorphisms $\diffeo^*_i, i=1, \cdots, n$ that align every surface $f_i$ onto the computed mean shape $\meansurf$:
\begin{equation}
	\left(\meansurf, \rotation^*_i, \diffeo^*_i \right)  = \argmin_{f, \rotation_i, \diffeo_i}  \sum_{i=1}^n d_{\preshapes}(f, \rotation_i f_i \circ \diffeo_i), 
	\label{eq:karcher_mean_registered}
\end{equation}  
where $d_{\preshapes}(\cdot, \cdot)$ is a certain measure of  dissimilarity (or distance) that quantifies shape differences. This equation can be simplified by writing it in terms of orbits or elements of the shape space $\shapesf$ as follows:
\begin{equation}
	[\meansurf]= \argmin_{[f] }  \sum_{i=1}^n d_{\shapes}([f],  [f_i]). 
	\label{eq:karcher_mean_shapespace}
\end{equation}  
where
$$
	d_{\shapes}([f_1],  [f_2]) = \min_{\rotation, \diffeo} d_{\preshapes}(f_1, \rotation f_2 \circ \diffeo).
$$

\noindent Similarly, one can define  the variance as  the expected value of the squared distance from the mean~\cite{frechet1948elements}:
\begin{equation}
	\sigma^2 = \frac{1}{n}\sum_{i=1}^n d_{\shapes}([\meansurf], [f_i])^2. 
\label{eq:non_linear_modes}
\end{equation}
Implementing this simple general formulation requires solving the following challenging fundamental problems;
\begin{itemize}
	\item \textbf{Choice of the metric $d_{\preshapes}$. } Computing shape statistics depends on the choice of the metric, or measure dissimilarity $d_{\preshapes}$. The challenge is in designing a metric that captures shape differences while being invariant to the set of transformations that preserve shape. 
	\item \textbf{The registration or correspondence problem. } Naturally, the solution to Eqn.~\eqref{eq:karcher_mean_shapespace} depends on the way the input surfaces are parameterized, or registered, \ie the process of finding the optimal rotations $O_i$ and diffeomorphisms $\diffeo_i$. 
	\item \textbf{Computational issues } that arise when using complex metrics that are physically motivated. One often deals with large collections of 3D models and thus the process of evaluating the metric, computing geodesics and shape statistics should be computationally efficient.
\end{itemize}
The following sections discuss these challenging problems and present different solutions that have been proposed in the literature.


\section{Shape spaces and metrics}

\label{sec:shape_spaces}

In what follows, we represent the boundary of a 3D object as an embedding $f: \stwo \to \real^3$. Here, we assume that the parameterization domain is the unit sphere.     It maps every point $s = (u, v) \in \stwo$ to a point $f(s) \in \real^3$.  For clarity purposes, we assume at this stage that the given surfaces have been normalized for translation and scale, thus they are elements of $\preshapesf$.


In shape analysis, a   dissimilarity measure quantifies  the amount of deformations, or energy, that one needs to apply to one shape in order to align it onto the other.  Consider the example of Fig.~\ref{fig:geod-syn} where $f_1$ is a straight cylinder and $f_2$ is a bended one. The deformation of $f_1$ onto $f_2$ results in a sequence of $m$ intermediate  shapes $F_i: \stwo \to \real^3, i=1, \dots, m$ such that $F_0 =  f_1,  F_m = f_1$, and $F_i = F_{i-1} + V_{i-1}$.
 Here, $V_{i-1}: \sphere \to \real^3$ is a (deformation) vector field such that $F_{i}(s) = F_{i-1}(s) + V_{i-1}(s)$.  The sequence $F = \{F_i, i=1,\cdots, m\}$ can be seen as a path or a curve in the preshape space $\preshapesf$. Its length is given by:
\begin{equation}
	L(F) = \sum_{i=1}^{m}  \sqrt{ \energy  \left( F_{i-1},  F_{i} \right)},
	\label{eq:discrete_path_length} 
\end{equation} 
where $ \energy $ is a certain measure of distance between the surfaces $F_{i-1}$ and $F_{i}$. When $m$ is sufficiently large, $F$ can be seen as a continuous deformation of $f_1$ onto $f_2$. It can be interpreted as a parameterized path $F: [0, 1] \to \surfaces$  such that $F(0) = f_1, F(1) = f_2$. Its length is given by:
\begin{equation}
	d_{\preshapes} (f_1, f_2) = L(F) = \int_0^1 \innerd{  \frac{ d F}{d\tau}(\tau)   }{\frac{d F}{d\tau}(\tau) }^{\frac{1}{2}} d\tau.
	\label{eq:continuous_path_length}
\end{equation}
Here, $ \frac{d F}{d\tau}(\tau) $ is in fact an infinitisemal  vector field that deforms the surface $F(\tau)$. It can be also interpreted as a small perturbation of the surface.  The inner product, $\innerd{\cdot}{\cdot}$, also known as \emph{the metric}, measures the strength  of this vector field.  Thus, integrating over $\tau$ provides the length of the path.

In general,  a surface $f$ is treated as an element, or a point, in a pre-shape space $\preshapesf$. A vector field that deforms $f$ is then a vector $\delta f$ that is tangent to $\preshapesf$ at $f$. Let $T_f(\preshapesf)$ denote the tangent space to $\preshapesf$ at $f$.  A metric is an inner product on $T_f(\preshapesf)$. It takes two tangent vectors $v_1$ and $v_2 \in T_f(\surfaces)$ and returns their inner product $\innerd{v_1}{v_2}$. The norm of a vector $v \in T_f(\surfaces) $, according to the metric $\innerd{\cdot}{\cdot}$, is given by $\innerd{v}{v}^{\frac{1}{2}}$. Since there are many paths that deform $f_1$ onto $f_2$, we are particularly interested in the shortest one, under the metric, which is called \emph{the geodesic} in the pre-shape space $\preshapesf$:
\begin{equation}
	F^* = \arg\min_FL(F).
	\label{eq:geodesic_path_optimization}
\end{equation} 
The length of the geodesic path is the dissimilarity, in the preshape space, between the two surfaces:
$$
	d_{\preshapes} (f_1, f_2) = \inf L(F) = L(F^*).
$$
Since almost every 3D shape analysis task includes a step in which shapes are compared based on some measure of dissimilarity, the choice of the metric is very critical for the subsequent analysis tasks.


\subsection{Kendall's shape space}
\label{sec:kendall}

The fundamentals  of  statistical analysis of shapes have been laid by Kendall~\cite{kendall:1977DS}, as early as $1977$, and advanced by many others~\cite{le1993riemannian,kendall2009shape,small1996statistical,dryden1998statistical}.  The basic idea is to   represent  the shape of a 3D object using  a finite set  $\textbf{p}$ of $\landmarkcount$ ordered points, called \emph{landmarks},  $\textbf{p} = \{p_i \in \mathbb{R}^3, i=1, \dots \landmarkcount \}$.  These points are sampled from the shape boundary and put in correspondence across shapes.  

In the case of our setup where a surface $f$ is treated as a mapping from a parameterization domain $\domain$ to $\real^3$, Kendall's representation is equivalent to having a fixed finite sampling of the parameterization domain $\domain$. That is $D = \{s_i, i=1, \dots, \landmarkcount \}$. As such, the space $\surfaces$ of all surfaces that can be represented with $\landmarkcount$ landmarks is $\real^{3\landmarkcount}$.  The pre-shape space, $\preshapesf$, \ie the space of surfaces that are normalized for translation and scale, is a hypersphere $\sphereone^{3\landmarkcount-4}$, which is a subspace of $\real^{3\landmarkcount}$.  To make the representation invariant to rotations, Kendall  defines the group action $SO(3) \times \sphereone^{3\landmarkcount-4}, (O, f) = Of$ whose orbits are  given by $[f] = \{(O, f) | O\in SO(3)\}$. The  shape space $\shapesf$ can then be viewed as  the quotient space $\sphereone^{3n-4} / SO(3)$.  The rotation-invariant geodesic distance between two surfaces $f_1$ and $f_2$ is given as the geodesic distance between their orbits on $\shapesf $, \ie:
\begin{equation}
	\small{d_\shapes([f_1], [f_2]) = \min_{O \in SO(3)} d_{\preshapes}(Of_1, f_2) = \min_{O \in SO(3)} d_{\preshapes}(f_1, Of_2),}
	\label{eq:kendall_dissimilarity}
\end{equation}
where $ d_{\preshapes}$ is the length of the arc on $\preshapesf$ that  connects $f_1$ to $Of_2$.  A geodesic path is then a great circle on $\preshapesf$   between  $f_1$ and $O^*f_2$ where
$
	O^{*} = \displaystyle\argmin_{O \in SO(3)} d_\preshapes(f_1, Of_2). 
$
Kendall's shape space,  introduced in 1977, and its subsequent developments~\cite{le1993riemannian,kendall2009shape,small1996statistical,dryden1998statistical}, are considered as the foundation that lead to the major recent developments in statistical shape analysis.

\subsubsection{Morphable models}
\label{sec:morphable_models}


In the case of planar objects, simply treating  Kendall's representation as a vector space, \ie assuming that the preshape space $\preshapesf$ is Euclidean, and $\innerd{\cdot}{\cdot}$ is  an $\ltwo$ metric,  results in the  2D  Active Shape Models (ASM)  of  Cootes et al.~\cite{Cootes1995ASM,cootes:eccv1998,cootes:pami2001}.  This formulation has  been later used for the analysis of the shape of 3D objects that are in static poses, leading to what is now known as  \emph{morphable models}. Morphable models have been  originally introduced for the analysis of the 3D shape of human faces~\cite{Blanz:1999:MMS}. They have been later used for the analysis of the 3D shape of  human bodies~\cite{Allen:2003} as well as the shape of objects originating from various domains including  archaeology, astronomy, morphometrics,  and medical diagnosis~\cite{dryden1998statistical,bhattacharya2008statistics,small1996statistical}.

Mathematically, for a given pair of vectors $v_1$ and $v_2 \in T_f(\preshapesf)$,  the $\ltwo$ metric is defined as:
\begin{equation}
	\innerd{v_1}{v_2} = \inner{v_1}{v_2},
	\label{eq:inner_L2_metric}
\end{equation}
where $\inner{\cdot}{\cdot}$ is the standard inner product. The geometric interpretation of this metric is as follows; A surface $f_1$ can be optimally deformed to another surface $f_2$ by simply adding to each point $f_1(s)$ a displacement vector $\disp(s)$ such that
$
	f_2(s) = f_1(s) + \disp(s).
$
The optimal path $F^*$ that connects $f_1$ to $f_2$, induced by the metric of Eqn.~\ref{eq:inner_L2_metric},  is simply a straight line. Subsequently, the difference, or dissimilarity,  between $f_1$ and $f_2$ is given by 
\begin{eqnarray}
	{ d(f_1, f_2) = \int_{\stwo} \|\disp\|ds  =   \int_{\stwo}  \|f_1(s) - f_2(s)\| ds.}
	\label{eq:L2metric}
\end{eqnarray} 
Assuming that $d(\cdot, \cdot)$ is Euclidean, and under a fixed parameterization, finding the optimal rotation that aligns one surface onto another is equivalent to finding the  rotation $\rotation\in SO(3) $ that minimizes $ \|f_1 - O f_2\|$.  
This   results in the Procrustean metric on the shape space $\shapesf$, which can be efficiently solved using Singular Value Decomposition (SVD)~\cite{goodall1991procrustes}. Now,  for a fixed rotation and reparameterization, \ie  the surfaces are already registered,  both the mean shape of Eqn.~\eqref{eq:karcher_mean_registered} as well as the covariance matrix $\cov$  have analytical solutions:
$$
\small{
	\meansurf = \frac{1}{\nshapes}\sum_{i=1}^{\nshapes} f_i,  \text{ and }
	\cov = \frac{1}{\nshapes-1} \sum_{i=1}^{\nshapes} (f - \meansurf)\times  (f - \meansurf)^T.
	}
$$
Thus, efficient parameterization of the shape variability in a collection of 3D shapes can  be performed using Principal Component Analysis (PCA). Let $\lambda_k, k=1, \dots, d$, be the leading eigenvalues of  $K$,  and $v_k$ their corresponding eigenvectors, called \emph{modes of variation}. A new statistically feasible shape $f$  is given by
$	f =  \meansurf   + \sum_{i=1}^{d} \alpha_k v_k,
	\label{eq:new_shape_pca}
$  where $\alpha_k \in \real$ are coefficients that control the contribution of each mode of variation.  Arbitrary new shapes can be generated by varying the coefficient vector  $\overrightarrow{\alpha} = (\alpha_1, \alpha_2, \dots, \alpha_d)$ whose  probability follows a Gaussian distribution  defined as
$	-\log{\text{Pr}(\overrightarrow{\alpha} )} = \frac{1}{2}  \sum_{i=1}^{d}\frac{ \alpha_i^2}{\lambda_i} + \text{const}
	\label{eq:linear_gauss_model}
$.  Thus, if one wants to synthesize a plausibe shape, \ie a shape that is similar, but not identical, to those in the collection, one only has to generate arbitrary $\alpha$s and retain those that have high probability.

This approach is known as \emph{morphable models} and is an  extension of the \emph{Active Shape Models} (ASM) developed by    Cootes et al.~\cite{Cootes1995ASM,cootes:eccv1998,cootes:pami2001} for the analysis of planar objects.
 They have been first introduced to 3D shape analysis by Blanz and Vetter in their seminal work on the synthesis of 3D faces~\cite{Blanz:1999:MMS}.  Morphable models have   been later generalized to characterize the space of the entire human body shapes~\cite{Allen:2003}.  Both in~\cite{Blanz:1999:MMS} and~\cite{Allen:2003}, the objects under study are assumed to be in neutral poses and their landmarks are in a one-to-one correspondence.


\subsubsection{The non-linear nature of Kendall's shape space}

\begin{figure}[t]
    \centering
   
    \begin{tabular}{@{}ccc@{}}
		 \includegraphics[width=.12\textwidth]{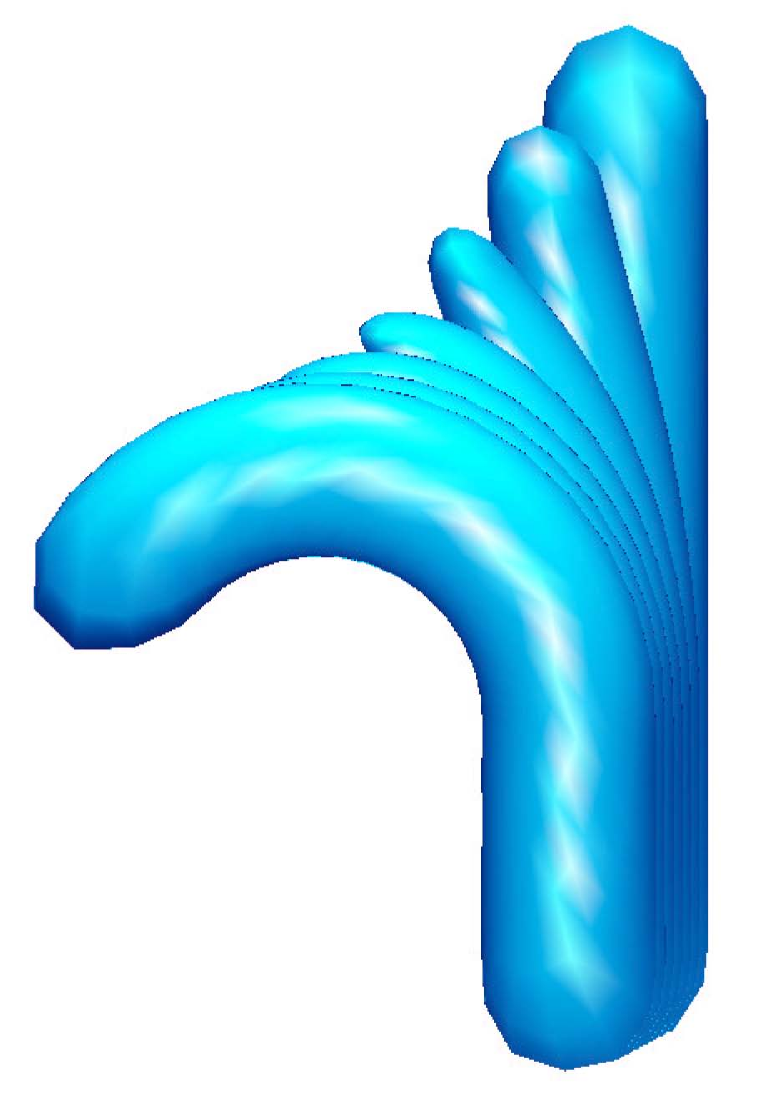}& 		
		  \includegraphics[width=.12\textwidth]{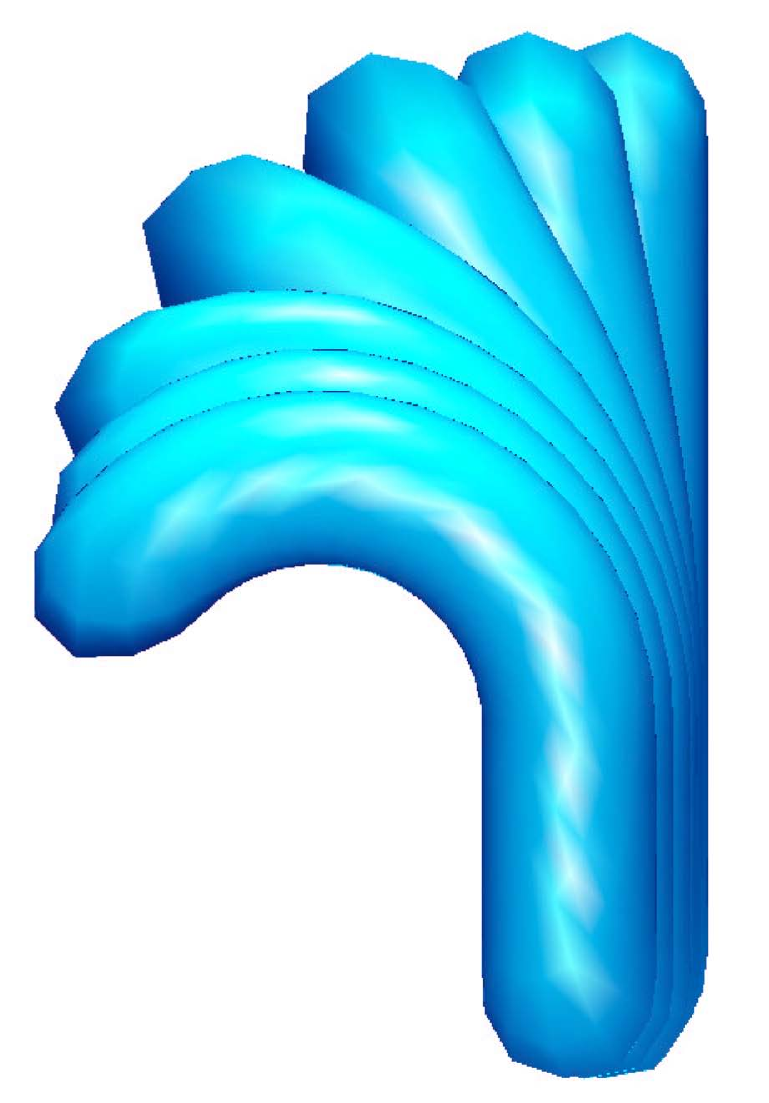}\\		 
		\small{(a) Linear path}    & \small{(b) Geodesic path}\\
		\small{$(1-\tau)f_1 + tf_2$}  & \small{by SRNF inversion }\\
    \end{tabular}
    \caption{\label{fig:geod-syn} Linear interpolation in $\mathcal{F}$ vs. geodesic path by SRNF inversion (see Section~\ref{sec:registration_geodeiscs}). Image courtesy of~\cite{laga2016numerical}.} 
\end{figure}


Although the $\ltwo$ metric on Kendall's shape space has been extensively used in a wide range of applications, \eg   archaeology, astronomy, morphometrics, medical diagnosis~\cite{dryden1998statistical,bhattacharya2008statistics,small1996statistical}, it is only suitable for computing geodesics, and thus statistics, between shapes that undergo small deformations, \eg  anatomical shapes,  3D faces, or human shapes in a neutral pose.   This main limitation is illustrated in Figure~\ref{fig:geod-syn} using two synthetic cylinders.  The example shows optimal deformation paths between surfaces $f_1$ and $f_2$, where $f_1$ is a straight cylinder and $f_2$ is a bent cylinder. Figure~\ref{fig:geod-syn}-(a) shows the linear path between $f_1$ and $f_2$ in $\shapesf$ (\ie after full registration) obtained using the $\ltwo$ metric by connecting  each pair of corresponding points with a straight line. The intermediate shapes along this path shrink unnaturally. Figure~\ref{fig:geod-syn}-(b) shows a natural  geodesic path computed with a metric that captures the shape deformations (here using SRNF inversion, which will be described later).

Kilian et al.~\cite{Kilian:2007:GMS}  build on Kendall's representation a new metric that does not suffer from the shrinkage issue of the $\ltwo$ metric. The idea is to define the distance between a surface $f_1$ and $f_2$ as the strength of a vector field $u \in T_{f_1}(\surfaces)$ that penalizes elastic deformations, enforcing as rigid-as-possible and as-isometric-as-possible deformations along geodesic paths.   That is, given a shape $f \in \surfaces = \mathbb{R}^{3n}$,  define  the inner product between two vectors $u$ and $v$ in $T_f(\surfaces)$  as:
\begin{equation}
\small{
 	\innerd{u}{v} =  \innerd{u}{v}^{R} + \innerd{u}{v}^{I} + \lambda\sum_{s=1}^{n} \inner{u(s)}{v(s)}a(s), 
 }
 \label{eq:kilian_metric}
\end{equation}
where $\lambda = 0.001$ and $a(s)$ is the local Voronoi area at the vertex $s$. The first term penalizes non-rigid deformations. To define it, consider first that the deformation vector $u$ can be written as $u= u' + u''$ where $u'$ is only composed of rigid transformations and $u''$ is only composed of non-rigid deformations. To have a metric that is invariant to rigid transformations while quantifying non-rigid ones then one can set $\innerd{u}{u}^{R}  =  \inner{u''}{u''}$. As such,  $ \innerd{u}{u}^{R} = 0$ if and only if $u$ is a rigid motion. This term  ensures invariance to translation, scale and rotation without normalization in  a pre-processing step.    

The second term of Eqn.~\eqref{eq:kilian_metric} favours bending over stretching.  It is defined as the strength of the deformations that remain after discarding isometric motions, \ie motions that preserve geodesic distances along the surface of $f$.  In other words, it can be seen as a measure of stretch in terms changes in the length of curves along the surface $f$. Since Kilian et al.~\cite{Kilian:2007:GMS}'s approach works on triangulated meshes then
$$
\innerd{u}{u}^I = \sum_{(s, t)} \inner{u(s) - u(t) }{ f(s) - f(t)  }^2, 
$$ 
where $s$ and $t$ are two adjacent vertices on the triangulated mesh that represents the surface $f$.   Thus, $ \innerd{u}{u}^{I} = 0$ if and only if  $u$ just bends the shape $f$. In other terms, when deforming the straight cylinder of Figure~\ref{fig:geod-syn}, the metric will favor bending over stretching and thus it will minimize the shrinkage that is observed when using the standard $\ltwo$ metric.

The last term is a regularizer that makes the metric Riemannian.  It also ensures that the geodesic  between two isometric shapes is unique.  The main limitation of this metric is the choice of the parameter $\lambda$ in order to  ensures that the metric is Riemannian. Large values of $\lambda$ will give high weight to the Euclidean term of Eqn.~\eqref{eq:kilian_metric}. Thus,  the metric may suffer  from the shrinkage issue observed in morphable models (see Figure~\ref{fig:geod-syn}). Note that Kilian et al. empirically set this parameter to $0.001$.


\subsection{Metrics that capture physical deformations}
\label{sec:elastic_metrics}

Instead of using the $\ltwo$ metric on $\preshapesf$, one would like to explicitly capture the type of deformations that one needs to apply to $f_1$ in order  to align it onto $f_2$. Such deformations can be of two types; bending and stretching.  This requires redefining Eqn.~\eqref{eq:continuous_path_length} in terms of an energy function that penalizes these deformations:
\begin{eqnarray}
	\label{eq:discrete_elastic_energy} 
	\small{ d_{\preshapes}^2(f_1, f_2)} &=& \small{\energy(f_1, f_2) }\\
	\nonumber		    		       &= & \small{ \int_0^1  \left\{ \alpha_s \stretchterm  \left( F(\tau) \right)   +    \alpha_b \bendterm  \left( F(\tau)  \right) \right\} d \tau},	
\end{eqnarray}
where $\energy_{\text{s}}$ and $\energy_{\text{b}}$ are respectively the stretching and bending energies. The parameters $\alpha_s$ and $\beta_s \in \realplus$ control the contribution of each of the terms to the energy function. Below we discuss two important pieces of work that implement this model. 

The first one (section~\ref{sec:thin_shells}) is inspired by the elasticity theory in physics~\cite{ciarlet2005introduction} where surfaces are treated as \emph{thin shells}, \ie a thin three-dimensional material of thickness $\delta$. Stretching in this case is caused by in-layer (tangential) shear or compression while bending is caused by friction due to transversal shear~\cite{Heeren:2012:TGS}. The second one (Section~\ref{sec:srnfs}) showed that with specific choices of the bending and stretching terms as well as their weights, the deformation model of Eqn.~\eqref{eq:discrete_elastic_energy} reduces to an $\ltwo$ metric in the space of Square Root Normal Fields (SRNF), a special representation of surfaces~\cite{Jermyn:2012:ESM}. 

\subsubsection{The shape space of thin shells}
\label{sec:thin_shells}

Based on a physical deformation paradigm, a surface $f$ can be viewed as the middle layer of a thin viscous material of some finite (albeit small) thickness $\delta$~\cite{zhang2015shell}.    Wirth et al.~\cite{Wirth:2011:CMA}, Heeren et al.~\cite{Heeren:2012:TGS,heeren2016splines},  Berkels et al.~\cite{BeFlHe13}, and Zhang et al.~\cite{zhang2015shell}  used the deformation model of Eqn.~\eqref{eq:discrete_elastic_energy}, with $\alpha_s = 1$ and $\alpha_b$ proportional to the squared thickness $\delta^2$. The stretch energy, referred to as \emph{the membrane} term, is given as $\stretchterm(f_1, f_2) = \int_{\stwo} \stretchterm(\cauchygreen(s)) ds$, where
	\begin{eqnarray}		
		\nonumber	\small{\stretchterm(\cauchygreen(s) ) }  &=&   \small{ \frac{\mu}{2} \trace(\cauchygreen(s)) + \frac{\lambda}{4}\det(\cauchygreen(s)) - } \\
	  				  &   &\small{ \left(  \frac{\mu}{2}  +  \frac{\lambda}{4}  \right)  \log \det  (\cauchygreen(s)) - \mu -\frac{\lambda}{4}.}
		\label{eq:trace_det_cauchygreen}
	\end{eqnarray}	
The $\log\det$ term penalizes material compression. $\lambda$ and $\mu$ are the Lam\'{e} constants of the tangential Newtonian dissipation measure.   The Cauchy-Green strain tensor,  $\cauchygreen =   \ffform_1^{-1} \ffform_2$, where $g_i$ is the first fundamental form or metric of the surface $f_i$,  accounts for changes in the metric  between the source and target surfaces.  In particular, the trace of the Cauchy-Green strain tensor, $\tr(G)$,  accounts for local changes in curve lengths while the determinant $\det(G)$ accounts for local changes in surface area~\cite{zhang2015shell}.  Thus,  this can be seen as the generalization of the stretch metric of Kilian et al.~\cite{Kilian:2007:GMS}, which only considered changes in the length of curves. 

The bending energy should account for out-of-plane bending and changes in curvature. Since such information is encoded in the shape operator, then the most natural choice of measure for quantifying bending is variation in the shape operator~\cite{Heeren:2012:TGS}:
\begin{equation}
	\bendterm  \left( f_1, f_2  \right) =  \int_{\stwo} \| \shapeop_2(s)) -  \shapeop_1(s) \|_F^2 ds,
	\label{eq:bending_energy_norm_shapeop}
\end{equation}
where $\|\cdot\|_F$ denotes the Frobenius norm. The energy of Eqn.~\eqref{eq:bending_energy_norm_shapeop} takes into account  full change in the second fundamental form.  Heeren et al.~\cite{heeren2016splines} and Windheuser et al.~\cite{windheuser2011geometrically} simplified this metric by only considering changes in the mean curvature, which is the trace of the shape operator $\shapeop$;
	\begin{equation}
		\bendterm  \left( f_1, f_2  \right) =  \int_{\stwo} \|\trace(\shapeop_2(s)) -  \trace(\shapeop_1(s)) \|^2 ds.
		\label{eq:bending_mean_curvature}
	\end{equation}
This energy is also known as the \emph{Willmore energy}. In the discrete setup, Heeren et al.~\cite{heeren2014exploring} approximated the mean curvature by the dihedral angle, \ie the angle between the normals of two adjacent faces on the triangulated mesh.

Since this deformation model acts on the Cauchy-Green tensor $\cauchygreen$ and mean curvature $\meancurvature$ then one can define the preshape space as the space of all pairs  $(\cauchygreen,  \meancurvature )$.   An important observation that has been made in~\cite{Heeren:2012:TGS,zhang2015shell,rumpf2014geometry} is that the Hessian of the elastic deformation energy results in a proper Riemannian metric on the space of shells, modulo rigid body motions. That is, for an infinitesimal deformation  $v \in T_{\preshapes}(f)$, $\Hess(\energy)(v, v) = D^2 \energy(v, v)$ is a proper metric that measures the strength of this deformation. (Here, $ D^2 \energy(v, v)$ is the second derivative of $\energy$ in the direction of $v$.)   Moreover,  Heeren et al.~\cite{Heeren:2012:TGS} showed that $\Hess(\energy)(v, v)  = 0$ if and only if $v$ induces an infinitesimal rigid motion. This last property is highly desirable in shape analysis where, as described in Section~\ref{sec:shape-preserving-transformations}, one seeks a framework that is invariant to rigid motions of shapes. This properties guarantees this invariance without normalization for translation and rotation of the shapes being analyzed. Shapes, however, need to be normalized for scale.

%


\subsubsection{The shape space of square-root representations}
\label{sec:srnfs}


Instead of using the Hessian of the deformation energy, Jermyn et al.~\cite{Jermyn:2012:ESM} showed that by representing a surface $f$ using the pair $(\ffform,  \unitnorm)$,  one can define a full elastic metric on the product space of metrics (\ie first fundamental forms) $\ffform$ and unit normal fields $\unitnorm$  as follows:
\begin{eqnarray}	
	\nonumber  \small{   \innerd{ (\delta \ffform, \delta \unitnorm)}{ (\delta g, \delta \unitnorm) } } = \small{  \int_{\stwo} ds  \sqrt{ |g| }   \{  \alpha \trace(  (g^{-1}  \delta   g)^2   )  + } \\
	 	  \small{  \beta \trace(g^{-1}  \delta   g )^2      +   \mu \inner{\delta \tilde{n}(s)}{\delta \tilde{n}(s) } 		  \}.}
		 \label{eq:jermyn_full_elastic_metric}
\end{eqnarray}
The first term of Eqn.~\eqref{eq:jermyn_full_elastic_metric}   accounts for changes in the shape of a local patch that preserve patch area. The second term quantifies changes in the area of the patches. Thus, the first two terms capture full stretch. The last term measures changes in the direction of the surface normal vectors and thus captures bending. The parameters  $\alpha, \beta$, and $\mu$ are positive weights that one can use to penalize the different  types of deformations, \eg  if one wants to favor stretch over bending then the third term should be given a higher weight.  As such, instead of working in the original space, \eg   $\preshapesf$ and $\shapesf$, one can define a mapping $\qmap$ such that $\qmap(f) = (\ffform, \unitnorm)$ where $\ffform(s)$ is the the first fundamental form of $f$ at $s$ and  $\unitnorm(s)$ is the unit normal vector to the surface at $s$.  The  preshape space $\preshapesq$ is then the product space of metrics and unit normal vectors. The shape space is the  
quotient space $\shapesq = \preshapesq /(SO(3)\times \diffeos)$, via the action of the rotation and re-parameterization groups.

Jermyn et al.~\cite{Jermyn:2012:ESM} defined a new representation of surfaces called the \emph{square-root normal fields} (SRNF), which are essentially   surface normals scaled by the square root of the local area: 
$$
	Q(f ) (s) = q(s) = \frac{\normal(s)}{\sqrt{ |  \normal(s)  | }  }
$$
where $\normal(s) = \frac{\partial f}{\partial u} \times \frac{\partial f} {\partial v}$ and $s = (u, v) \in \sphere$. The space of SRNFs, hereinafter denoted by $\srnfs$, has very nice properties that are relevant for shape analysis. In particular, Jermyn et al.~\cite{Jermyn:2012:ESM} showed that the  $\ltwo$ metric on $\srnfs$ is a special case of the full elastic metric given by  Eqn.~\eqref{eq:jermyn_full_elastic_metric}. It corresponds to the case where $\alpha = 0, \beta = \frac{1}{4}$, and $\mu = 1$:
\begin{equation}
\small{
	\inner{\delta q}{\delta q} =  \int_{\stwo} ds  \sqrt{ |g| }   \{   \frac{1}{4} \trace(g^{-1}  \delta   g )^2      +    \inner{\delta \tilde{n}(s)}{\delta \tilde{n}(s)} \}.
	}\label{eq:srnf_metric1}
\end{equation}
Since $ \sqrt{ |g| }  $ is just $|n|$, the Eqn.~\eqref{eq:srnf_metric1} is equivalent to
\begin{equation}
\small{
	\inner{\delta q}{\delta q} =    \frac{1}{4}  \int_{\stwo} ds \{  \frac{\delta a(s) \delta a(s)}{ a(s) }         +    \ \inner{\delta \tilde{n}(s)}{\delta \tilde{n}(s)} a(s)   \}.
	}
\end{equation}
Thus, the $\ltwo$ metric in the space of SRNFs is equivalent to a weighted sum of surface area change and surface bending. This property of SRNFs makes them very promising as a representation of surfaces for elastic shape analysis.  If geodesics, mean shapes, PCA, etc.\ could be computed in $\srnfs$ under the $\ltwo$ metric, and then mapped back to $\surfaces$ or $\preshapesf$ then there would be large gains in computational efficiency with respect to other metrics such as those defined on the space of thin-shells. 
 Unfortunately, there is no analytical expression for $Q^{-1}$ for arbitrary points in ${\srnfs}$. Moreover, the injectivity and surjectivity of $Q$ remain to be determined, meaning that for a given $q\in\srnfs$, there may be no $f\in \preshapesf$  such that $Q(f) = q$, and if such an $f$ does exist, it may not be unique.
 
If one cannot invert the representation, one can always pull the $\ltwo$ metric back to $\preshapesf$ under $Q$ and perform computations there, as in ~\cite{xie-iccv:2013}, but this is computationally expensive, and rather defeats the purpose of having an $\ltwo$ metric in the first place.  Although an analytical solution to the inversion problem remains an open problem, Laga et al.~\cite{laga2016numerical} showed that one can always estimate, for a given $q \in \srnfs$ a surface $f\in \preshapesf$ whose SRNF representation is as close as possible $q$.

SRNF representation is a generalization of the Square-Root Velocity Function (SRVF)~\cite{srivastava2011shape} proposed for elastic analysis of planar curves and used in many applications including biology and botany~\cite{laga2012riemannian,laga2014landmark}. It is also  a special case of the family of square-root maps (SRM). Another example of SRM is the Q-maps introduced  by Kurtek et al.~\cite{kurtek2010novel,kurtek:mi2011,kurtekipmi2011,kurtek:pami2012,kurtek_mmbia2012}, which, similar to SRNF maps,  is used for the analysis of the shapes of parameterized surfaces using the $\ltwo$ metric.  It is defined a mapping $Q: \surfaces \to \ltwo$, where $q = Q(f)$ such that  $q(s) = \sqrt{|n(s)|}f(s)$.

\vspace{6pt}\noindent\textbf{Properties. }
SRNF and Q-maps share very nice mathematical properties that make them suitable for shape analysis. In particular, both SRNFs and Q-maps are a subset of $\mathbb{L}^2(\mathbb{S}^2, \mathbb{R}^3)$. This means that one can use the standard $\ltwo$ metric in the space of SRNFs or Q-maps for the analysis of shapes instead of the complex and expensive pullback elastic metrics. Furthermore, if a surface is re-parameterized according to $f \to f \circ \gamma, \gamma \in \Gamma$,  its corresponding Q-map or SRNF is given by $(q, \gamma) = (q\circ\gamma)\sqrt{J_{\gamma}}$, where $J_{\gamma}$ is the determinant of the Jacobian of $\gamma$.

Another  main motivation  behind the use of  Q-maps and SRNFs in surface analysis is that the action of $\Gamma$  is by isometries under the $\ltwo$ metric, \ie $\|q_1 - q_2\| = \|( q_1, \gamma) - ( q_2, \gamma)\|$,  $\forall \gamma \in \Gamma$.  This property is very important for joint rotation and re-parameterization-invariant comparison, and thus elastic registration, of surfaces as will be discussed in Section~\ref{sec:registration_geodeiscs}. As such, Q-maps and SRNF maps have been used for optimal re-parameterization, and thus registration,  of spherical and quadrilateral surfaces~\cite{kurtekipmi2011,kurtek:mi2011,Jermyn:2012:ESM,laga2016numerical} that undergo isometric as well as elastic deformations,  and surfaces that contain missing parts. They have been also used for computing geodesic paths between  such surfaces~\cite{kurtek:pami2012,laga2016numerical}. The strength of these two approaches is that shape comparison, registration and geodesic computations  are performed under the same elastic Riemannian metric.  

The main difference between Q-maps and SRNFs is that (1) the former lacks a physical interpretation;  it does not have a clear relationship to an underlying elastic metric. The representation was solely devised for convenience of being able to compare the shapes of parameterized surfaces using the $\ltwo$ metric, and (2) unlike Q-maps, SRNF representation is invariant to translation since it is computed using the surface normals.

\subsection{Transformation-based representations}

\label{sec:transformation-based representations}

The approaches described in Sections~\ref{sec:kendall}  and~\ref{sec:elastic_metrics} represent a shape as an embedding $f: \domain \to \real^3$ while the metric quantifies the deformations the shape undergoes. Another approach is to model transformations that 3D objects undergo rather than the transformed objects themselves. These representations derive from Grenander's pattern theory~\cite{Grenander:1998:CAE} in which a shape is not represented as such but as a deformation of  another (fixed) shape, $T$,  called template~\cite{Younes:survey2012}. This paradigm has been introduced by Grenander~\cite{Grenander:1998:CAE}, developed in~\cite{beg2005computing,glaunes2008large,vialard2012diffeomorphic,vialard2012atlas} and applied to various problems in computational anatomy and more recently to the analysis of human body shapes~\cite{Anguelov:2005}.

Let's assume that we are given a set of registered surfaces represented as triangulated meshes with the same graph topology as the template $T$.  We assume that the template has $m$ triangular faces, $T=\{T_k, k=1\dots m\}$. Each  triangular face can be represented with a matrix of its three vertices,  $[v_0, v_1, v_2] \in \mathbb{R}^{3\times3}$, or with its edge matrix $\textbf{e}=[v_1 - v_0, v_2 - v_0] \in \mathbb{R}^{3\times2}$.  Any surface   $X_i = \{X^i_k, k=1\dots m\}$ of  $m$ triangles  can be represented with a set of transformation matrices $P_i = \{P^{i}_{k}, k=1\dots m \}$  that deform the template $T$, \ie $X^i_k = P^i_k T_k$ where    $P^i_k$  is a ${3\times3}$  matrix that encodes various types of deformations that  $T^i_k$ can undergo.  Hereinafter, we refer to  $P$ as the \emph{deformation model}.

Existing methods  differ in the choice of (1) the  template $T$,   (2) the deformation model $P$, and (3) the metric used to measure distances in the space of deformations.   We will discuss these aspects in the following subsections.

\subsubsection{Choice of the template $T$ } 

When dealing with 3D models of the same class, \eg  human body shapes, one can define the template $T$ as a typical 3D model in a neutral pose. This has been used in the SCAPE model~\cite{Anguelov:2005} and in many other papers~\cite{DBLP:journals/cgf/HaslerSSRS09,Hasler:2009:TSE} for the analysis of human body shapes.  In general, the template $T$ can be any surface, \eg  a sphere, of same topology and triangulation as the surfaces being analyzed.  In practice, however, domain specific templates are desirable to improve robustness and computation time. Alternatively,   one can define $T$ as a single canonical triangle in the $xy$-plane~\cite{DBLP:conf/cvpr/HaslerARTS10}.  An exception to these  representations is the SCAPE model~\cite{Anguelov:2005} where, in addition to  the template mesh $T$, an articulated skeleton is used to capture articulated motion.  These approaches operate on  meshes that are in full correspondence with exactly the  same connectivity.

\subsubsection{Deformation models}

The deformation model $P$ encodes  different deformations that a triangular face $X^{i}_{k} \in X_i$ undergoes with respect   to its corresponding triangle $T_k$ on $T$.  These  can be  rotations due to articulated motion (\eg  bending an arm of a human body), stretching due to shape differences between say two individuals in the same pose, or pose induced deformations (\eg  muscle stretch due to arm  bending).

 SCAPE~\cite{Anguelov:2005} used two deformation models, one acting on a template mesh and another on an articulated skeleton (which captures articulated motion). This  greatly simplifies the mathematical formulation;   A given mesh  $X_i$ is represented with  three sets of transformation matrices: $Q^i = \{Q^i_k \}$, $S^i = \{S^i_k \}$, and $R^i = \{R^i_{ l[k]}\}$  for  pose induced deformations,  body shape deformations, and articulated motions, respectively.   
  Here,  $R^i$ acts on an articulated skeleton used to capture articulated motion where $R^i_{ l[k]}$ refers to the rotation of the $l$-th part of the skeleton to which the $k$-th face belongs to. 
  
Using  separable deformation models assumes that intrinsic  shape deformations and  pose variations are independent. This, however,   prevents one  from capturing phenomena where there is a strong correlation between body pose and local deformations (\eg  near the joints or at the muscles). For example, the  surface deformation generated by the motion performed by an athletic person exhibits different properties than the same motion carried out by a person with less pronounced  muscles~\cite{DBLP:journals/cgf/HaslerSSRS09}. Separable models cannot capture such  correlation between pose and body shape.

To overcome this limitation, Hasler et al.~\cite{DBLP:journals/cgf/HaslerSSRS09,DBLP:conf/cvpr/HaslerARTS10}  introduced a bi-linear model of pose and shape, without using a skeleton, where   each triangular face $X^i_k$ is defined as $X^i_k = R^i_k S^i_k T$. The matrices $R^i_k$ and $S^i_k$, which represent the pose and shape components,  respectively, are affine transforms applied to the template  $T$.  Note that, while  SCAPE  requires a set of 3D shapes of the same  subject  in different poses to learn the space of pose deformations, Hasler et al.~\cite{DBLP:journals/cgf/HaslerSSRS09,DBLP:conf/cvpr/HaslerARTS10}  learn   both spaces using  a random mix of 3D models from different subjects  in various poses.  Both  methods have shown detailed  representation of the human body shape. However, using strong priors on a human body model, \ie  manually defined skeleton and body parts, makes these methods  difficult to extend to other, especially free-form, object categories without redesigning the representation~\cite{DBLP:conf/cvpr/HaslerARTS10}.

Another property of the representations above is that they have redundant degrees of freedom (DOF). For instance, Hasler et al. model deformations with $15$ DoF and a non-linear encoding of triangle deformations that captures dependencies between pose and
shape.  Freifeld and  Black~\cite{Freifeld:2012:LBM} showed that triangle deformations lie in a $6$ dimensional nonlinear manifold.   More precisely, $X_k^i = R^{i}_k O^{-1}_k A^{i}_k S^{i}_k  O_k T_k$, where $O_k$ is a rotation that maps the template triangle $T_k$ to a canonical planar triangle. The scaling  $S^{i}_k$ and transformation $A^{i}_k$ are in-plane transformations. Finally,  $R^{i}_k$ and  $O^{-1}_k$ are rotations that map the planar triangle to the 3D space. Note that $O_k$ is fixed since it depends on the template.  Freifeld and  Black also showed that $R^{i}_k O^{-1}_k A^{i}_k S^{i}_k  O_k$ is invertible and has $6$ DoF, which eliminates deformations that do not have physical meaning. A triangular mesh of $m$ faces is  then represented with $m$-tuples $(R, A, S)$, where $R\in SO(3)$, $A$ is a $2\times2$ matrix, and $S>0$. 


%

Finally, medial representations (M-reps), which are an extension of curve skeletons~\cite{sundar2003skeleton} and medial surfaces~\cite{Siddiqi:2008:RAM},   have been  introduced by Blum~\cite{blum1967transformation} as a tool  for representing the geometry and structure of anatomical shapes~\cite{pizer1999segmentation,joshi:202:medial,Fletcher:2003:SSV,fletcher2004principal}.   Unlike the deformation models described above, which act on the shape's surface, a medial surface is formed by the centers of all spheres that are interior to the object and tangent to the object boundary in two or more points.
Each medial point, called \emph{medial atom},   is a 4-tuple $\textbf{m} = \{\textbf{x}, r, \textbf{n}_0, \textbf{n}_1\}$, where $\textbf{x} \in \mathbb{R}^3$ is the location of the atom, $r \in \mathbb{R}^+$ is the radius of the  maximum sphere that is tangent to the surface in two or more points, $\textbf{n}_0$ and $\textbf{n}_1 \in \mathbb{S}^2$ are two unit spoke directions.  A medial atom $\textbf{m}$  is an element of  the manifold $\mathcal{F}(1) = \mathbb{R}^3 \times \mathbb{R}^+ \times \mathbb{S}^2 \times \mathbb{S}^2$.  Fletcher et al.~\cite{fletcher2004principal} showed that by representing the atoms with the deformation, \ie translation, scaling, and rotation, of a canonical atom, \eg  $(0,0,1)$, they become elements of $\mathcal{S}(1) = \mathbb{R}^3 \times \mathbb{R}^+ \times SO(3)/SO(2)  \times SO(3)/SO(2)  $, which is a Lie group. A grid of $\landmarkcount$ medial atoms representing a 3D object is   an element of the Lie group $\mathcal{S} = \mathcal{S}(\landmarkcount)$, the product of $\landmarkcount$ copies of $\mathcal{S}(1)$.

Since their introduction, M-reps have been  used   in various medical image processing tasks, including segmentation~\cite{pizer2005method,pizer:2003ijcv}, registration, shape discrimination~\cite{bouix2005hippocampal}, and  brain-structure analysis~\cite{yushkevich2009continuous,gorczowski2010multi}.


\subsubsection{Metrics on the space of deformations }  
\label{sec:metrics_space_defo}
Representing shapes with the deformations of a template leads to shape spaces that are non Euclidean  (rotations, for example, are elements of $SO(3)$).  Some methods   typically use a Euclidean representation of deformations and measure distances  using the $\ltwo$ metric,  ignoring the geometry of the space of deformations. The SCAPE model~\cite{Anguelov:2005}, for instance, learned a shape model using the Euclidean distance and PCA on the  deformation matrices $S$ using only  the subjects that are in the same neutral pose. Similarly, Hasler et al.~\cite{Anguelov:2005} learned, separately,  
two low-dimensional models of  shape and pose, and then combined them at a later stage. 

Unlike these works, both M-reps~\cite{pizer1999segmentation,joshi:202:medial,Fletcher:2003:SSV,fletcher2004principal} and Lie bodies~\cite{Freifeld:2012:LBM} exploit the manifold structure of the space of deformations where geodesics and statistics are computed using proper Riemannian metrics. An important property that makes these methods of practical interest is that, despite the non-linearity of the metric as well as the manifold,  geodesic distances and geodesic paths between a pair of shapes have a closed-form  formula, unlike other methods which require expensive optimization schemes, see~\cite{Freifeld:2012:LBM} and~\cite{Fletcher:2003:SSV,fletcher2004principal} for the mathematical details.   This significantly simplifies the process of comparing shapes and computing geodesics and  summary statistics.

Note that these representations require   that all the shapes being analyzed to have the same connectivity and to be in one-to-one correspondence with each other and with the template.  Such correspondences need to be computed using a different approach, and more importantly, using an optimality criteria that  is often different from the one used for computing geodesics and statistics.

\begin{landscape}

\begin{table}
	\caption{\label{tab:taxonomy_equations} Taxonomy of the most representative papers where: $\landmarkcount$: number of landmarks or vertices, $\cauchygreen$: Cauchy-Green tensor, $\shapeop$: the shape operator, Trans.: translation, Rot.: rotation, Reg.: registration, $\preshapesf$: prehsape space of zero-centered and unit-area normalized surfaces, $\shapesf$: shape space, $\preshapesq$ preshape space of Q-maps or SRNFs, and $\shapesq$: shape space of Q-maps or SRNFs.
	}

	\begin{tabular}{ @{}p{.15\textwidth}@{ } p{.12\textwidth}  @{ }c @{ }c@{  }c @ { }c   p{.3\textwidth} @{ }  p{.1\textwidth} @{ } p{.45\textwidth}@{} }
	\toprule
	
	 \multirow{2}{*}{Methods} &  \multirow{2}{*}{Input} & \multicolumn{4}{c}{Normalization requirement }  &  \multirow{2}{*}{Preshape space}   &  \multirow{2}{*}{Metric} & \multirow{2}{*}{Geodesic distance $\energy(f_1, f_2) $}\\
	 	\cmidrule{3-6}
	 					& 				     & Trans. & Scale & Rot. & Reg. & &&  \\	
	\midrule
	Morphable models~\cite{Blanz:1999:MMS,Allen:2003}   & $f \in \real^{3\landmarkcount}$ &  \yes & \yes & \yes & \yes &  
	$ \mathbb{S}^{3\landmarkcount-4}$  &  $\ltwo$  & $\|f_1 - f_2\|$ \\
	
	\midrule

	
	Kilian et al.~\cite{Kilian:2007:GMS} & $f \in \real^{3\landmarkcount}$  & \no & \no &\no & \yes & $ \real^{3\landmarkcount}$ &   
		
		 \multicolumn{2}{@{}p{0.55\textwidth} }{\small{$\innerd{u}{v} =   \inner{u"}{v"} + 
	          \sum_{(s, t)} \inner{u(s) - u(t) }{ f(s) - f(t)  }   \inner{v(s) - v(t) }{ f(s) - f(t)  }   $
	          $+ \lambda\sum_{s=1}^{\landmarkcount} \inner{u(s)}{v(s)}a(s)$. $u"$ and $v''$ are the non-rigid component of $u$ and $v$, respectively.  
	          $\lambda=0.001$,  and $a(s) \text{ the Voronoi area at } s$  }}   \\

	\midrule
	
	Thin shells~\cite{Heeren:2012:TGS,heeren2016splines,BeFlHe13,zhang2015shell} & $f \in \real^{3\landmarkcount}$  & \no & \no & \no & \yes &  $\real^{3\landmarkcount}$  & 	$\Hess(\energy)$ &  
	$ \small{  \int_{\sphere}ds \{ \frac{\mu}{2} \trace(\cauchygreen(s)) + \frac{\lambda}{4}\det(\cauchygreen(s)) -   \left(  \frac{\mu}{2}  +  \frac{\lambda}{4}  \right)  \log \det  (\cauchygreen(s))  }  $ \\
	& & & & & & & 	&
	$ \small{ - \mu -\frac{\lambda}{4} + \|\trace(\shapeop_2(s)) -  \trace(\shapeop_1(s)) \|^2 \}.} $\\
	
	\midrule
	Q-maps\cite{kurtek2010novel,kurtek:mi2011,kurtekipmi2011,kurtek:pami2012,kurtek_mmbia2012}  & $f: \domain \to \real^3$,  $\domain = \stwo, [0, 1]^2$  &   \yes & \yes & \no & \no &  \small{ $\preshapesq = \{q = Q(f) | f \in \preshapesf; $}    &  $\ltwo$ on $\preshapesq$ &   \small{$\displaystyle  \inf_{O\in SO(3), \diffeo \in \diffeos}\| q_1 -O (q_2, \diffeo)\|,   q_i = Q(f_i)$  } \\
	  &  & & & &  & $ \small{q(s) = \sqrt{\|n(s) \|} f(s) \} }$  &   & \\
	
	\midrule
	SRNFs~\cite{Jermyn:2012:ESM,xie-iccv:2013,xie2014numerical,laga2016numerical}    & $f: \domain \to \real^3$, $\domain = \stwo, [0, 1]^2$  &   \no & \yes & \no & \no &  \small{ $\preshapesq = \{q = Q(f) | f \in \preshapesf; $}    &  $\ltwo$ on $\preshapesq$ &   \small{$\displaystyle  \inf_{O\in SO(3), \diffeo \in \diffeos}\| q_1 -O (q_2, \diffeo)\|, q_i = Q(f_i)$ }\   \\
	  &   & & & &  & $ \small{q(s) = \frac{n(s)} { \sqrt{\|n(s) \|}} \} } $  &   & \\ 
	
	\midrule
	
	Lie bodies~\cite{Freifeld:2012:LBM} &  $f \in \real^{3\landmarkcount}$    &  \yes & \yes & \yes & \yes &    \small{$(SO(3)\times G_A \times \real^+)^\landmarkcount$} &   &  \small{$ \sum_{j=1}^{\landmarkcount} d(g_{1,j}, g_{2,j}), g_i = (R_i, A_i, S_i)$}, \\
								& &&&& &  \small{$G_A = \left( \begin{array}{cc}
																		1 & u\\
																		0 & v
																		\end{array}\right), u \in \mathbb{R}, v > 0$}& &   $ d(g_1, g_2) =  \|\log(R_1^{-1} R_2) \|_F + \|\log(A_1^{-1} A_2) \|_F +  \|\log(S_1 / S_2) \|$, $R_i\in SO(3), A_i \in G_A, S_i \in \mathbb{R}^+$   \\

	\midrule
	M-reps~\cite{pizer1999segmentation,joshi:202:medial,Fletcher:2003:SSV,fletcher2004principal}  &  $f \in \real^{3\landmarkcount}$    &  \yes & \yes & \yes & \yes &    \small{ $\left(\real^3 \times \real^+ \times  \sphere  \times \sphere\right)^\landmarkcount $}  & & 
	\small{ $\sum_{i=1}^\landmarkcount   \|\text{Log}_{\text{m}_{i1}} (\text{m}_{i2}) \|$,   $\text{m}_{ik}:  i-$the medial atom of the $k-$th shape. $ \text{m}_{ik} \in  \real^3 \times \real^+ \times  \sphere  \times \sphere.$ }\\

	\bottomrule
	
	\end{tabular}

\end{table}

\end{landscape}

\section{Registration and geodesics}
\label{sec:registration_geodeiscs}

Now that we have introduced the concepts of metrics, pre-shape spaces, and shape spaces, we turn our attention to the problem of computing geodesics, \ie smooth optimal deformations, $F^*$,  between a source surface $f_1$ and a target surface $f_2 \in \preshapesf$:
\begin{equation}
	F^* = \min_{(O, \diffeo) \in SO(3)\times\diffeos}  \left\{ \min_{  \substack{{F:[0, 1] \to \preshapesf,} \\ {F(0) = f_1, F(1) = O(f_2 \circ \diffeo)}} } L(F)  \right\}, 
	\label{eq:geodesic_correspondence}
\end{equation}
where $L(F)$ is given by Eqn.~\eqref{eq:continuous_path_length} and depends on the choice of representation and metric. This double optimization problem finds the shortest path, and thus the geodesic, in the shape space between $[f_1]$ and $[f_2]$. The inner minimization problem assumes that the surfaces are in correspondence, \ie fixes the rotation $O$ and reparameterization $\gamma$, and finds the shortest path in $\preshapesf$ that connects $f_1$ to $O(f_2 \circ \diffeo)$.  The outer optimization is a registration step, which finds the optimal rotation  $O$ and reparameterization $\diffeo$ that bring  $f_2$   as close as possible to $f_1$ without changing its shape. 
Clearly, these two problems, \ie registration and geodesics, are interrelated and should be solved jointly using the same metric (or optimality criteria).



\subsection{Registration}

In shape analysis, registration quality is a major factor in determining the accuracy of the subsequent analysis steps such as computing optimal deformations (geodesics) or  summary statistics (\eg  mean shapes and modes of variations).  It has been extensively studied in the past two decades, see~\cite{vankaick11correspsurvey} and~\cite{tam2013registration} for comprehensive surveys and different taxonomies of the state-of-the-art. In general, the registration problem can be mathematically stated as the problem of finding the optimal  transformation, $T$, and a re-parameterization $\gamma: \sphere  \to \sphere$ such that $T (f_2 \circ \gamma)$ is as close as possible to $f_1$, \ie     $\int d^2\left(f_1(s),   T f_2(\gamma(s) \right) ds  $ is minimized. Here  $d(\cdot, \cdot)$ is a certain measure of closeness in the space of surfaces.  We have seen in Section~\ref{sec:shape-preserving-transformations} how to remove the variabilities due to translation and scale. In this section, we will focus on the remaining variables, which are rotations and reparameterizations. Thus, we are given two surfaces $f_1$ and $f_2 \in \preshapesf$ and we seek to solve the following optimization problem:
\begin{equation}
{
	(\rotation^*, \diffeo^*) = \argmin_{(\rotation, \diffeo) \in \rotations \times \diffeos }  d_{\preshapes} \left(f_1,   \rotation (f_2 \circ \diffeo) \right),
}
	\label{eq:registration_problem}
\end{equation}

\noindent The optimization problem of Eqn.~\eqref{eq:registration_problem} is a complex problem that can be solved by iterating between the optimization over rotations $O\in SO(3)$, \ie finding the optimal rotation, given known correspondences (\ie fixed $\diffeo$), and then finding the optimal correspondences or reparameterization $\diffeo$ given fixed rotation $O$. 

If correspondences are given,  finding the best  $\rotation$ that aligns $f_2$ onto $f_1$  is straightforward. Under the $\ltwo$ metric,  the best rotation $\rotation$ can be analytically found using Singular Value Decomposition (SVD) leading to the Procrustes analysis of shapes~\cite{gower2004procrustes}, which forms the basis of  2D Active Shape Models (ASM)~\cite{Cootes1995ASM} and  3D morphable models~\cite{Blanz:1999:MMS}. Similarly, if one is using the partial elastic metric of Jermyn et al.~\cite{Jermyn:2012:ESM} (Section~\ref{sec:srnfs}), then one can find the optimal rotation $O$ in the space of SRNFs and apply it to the original surface since the SRNF map of $Of$ is exactly $Oq$.

In practice, correspondences are   unknown and better results are obtained if  one  solves simultaneously for the best correspondences and for the optimal rigid  alignment.


\subsubsection{Landmark-based elastic registration}

In this class of methods, 3D shapes are treated as sets of landmarks sampled from their surfaces. The problem of finding correspondences between two surfaces $f_1$ and $f_2$ is then reduced to the problem of matching landmarks across the surfaces using some optimality criteria.
The sampling of landmarks can be either uniform or adaptive to cope with the varying surface complexity~\cite{osada2002shape,bronstein2008numerical,sohel2014survey}. For instance, in the case of triangulated meshes, each vertex can be treated as a landmark. In the case of parameterized surfaces, $f:\sphere \to \real^3$, then one can sample points $s_i$ from the domain $\sphere$  and use $f(s_i) = f_i$ as landmarks. 

A popular approach for putting in correspondence the landmarks is the Iterative Closest Point (ICP) algorithm, proposed in~\cite{besl_icp1992,chen1992object},  which iterates between two steps: (1) a matching step where correspondences are estimated either by nearest-neighbor search  using  point-to-point~\cite{besl_icp1992} or point-to-plane~\cite{chen1992object} distances, or using some local descriptors, and (2) an optimal alignment step using the estimated correspondences. Since  its introduction,  many  variants of the ICP algorithm have been proposed. They aimed at improving various aspects, such as  speed and quality of  convergence,   of the original algorithm, by, for example, combining multiple distance measures  with some local descriptors~\cite{rusinkiewicz2001efficient}. 
Nevertheless, ICP algorithm and its variants provide good results when the poses of the shapes being aligned are close to each other. 
When the difference in  pose  is large,  the search space for  correspondences is very large. In theory, however,  three pairs of corresponding points are sufficient to define a rigid transformation. This fact has been  used in~\cite{chen1999ransac,papazov2011stochastic,rodola2013scale} to derive RANSAC-based solutions. Their complexity is, however, high, of order $O(\landmarkcount^3)$, where $\landmarkcount$ is the number of points being aligned. Aiger et al.~\cite{aiger20084} introduced the four point congruent sets (4PCS) algorithm, which reduced the complexity of RANSAC algorithms to $O(\landmarkcount^2)$. It has been later optimized to achieve a complexity of order $O(\landmarkcount)$~\cite{mellado2014super}.

Another approach for finding correspondences is to characterize the shape's geometry around the landmarks using some descriptors~\cite{johnson1999using,Laga:2006:SWD,hedi:cvpr2014,tabia2015covariance,tabia2013compact,laga2007discriminative,laga20093d} and then match each landmark to its closest one in terms of a distance measure in the descriptor space~\cite{laga2008supervised,hamid2008supervised}. The main difficulty, however, is that  in the vast majority of applications, one is required to put in correspondence  two or more 3D objects that undergo non-rigid deformations that affect the geometry of the local regions. When restricting the analysis to surfaces that only undergo isometric deformations, \ie bending,  which is a subclass of non-rigid deformations, one  can design methods or descriptors that only capture the intrinsic properties of shapes, \ie the properties that remain unchanged  after changes due to bending are factored out. Examples of such approaches include Generalized Multidimensional Scaling (GMDS)~\cite{bronstein2006generalized}, spectral analysis based on Laplace-Beltrami eigenfunctions~\cite{rustamov2007laplace,zhang2010spectral}, and Heat Kernel Signatures (HKS)~\cite{sun2009concise,Bronstein:2010:GFD}.  Several  papers have attempted to relax the isometry constraint so that 3D objects can be put in correspondence even when they undergo stretching. Zhang et al.~\cite{zhang2008deformation}, for example, put two objects  in correspondence by finding the optimal bending and stretching that is needed to align one shape onto another. In this work, bending is measured in terms of changes in the direction of the normal vectors while stretching is measured in terms of changes in edge lengths. Thus, the approach uses a metric that is similar to those presented in Section~\ref{sec:shape_spaces}.

\begin{figure}[tb]
\centering{
\begin{tabular}{@{}c@{}c@{}}
			\includegraphics[width=.25\textwidth]{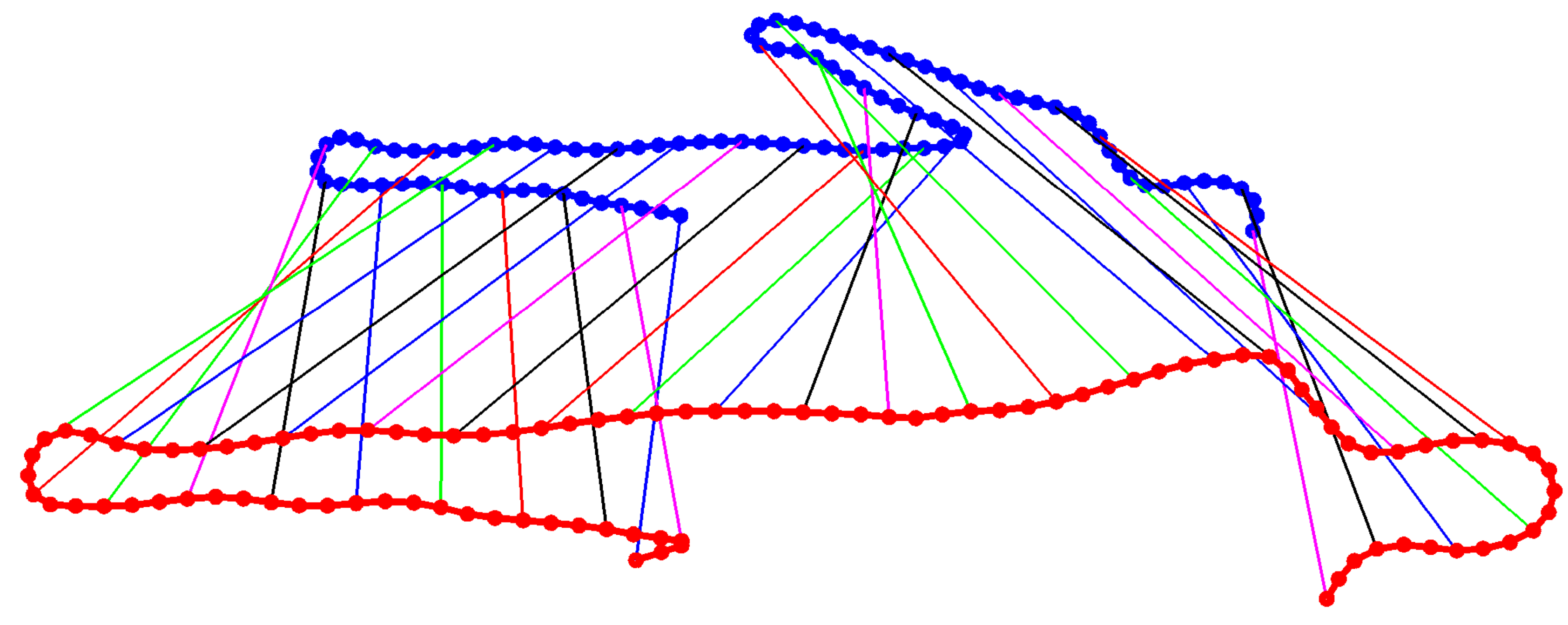}&
			\includegraphics[width=.25\textwidth]{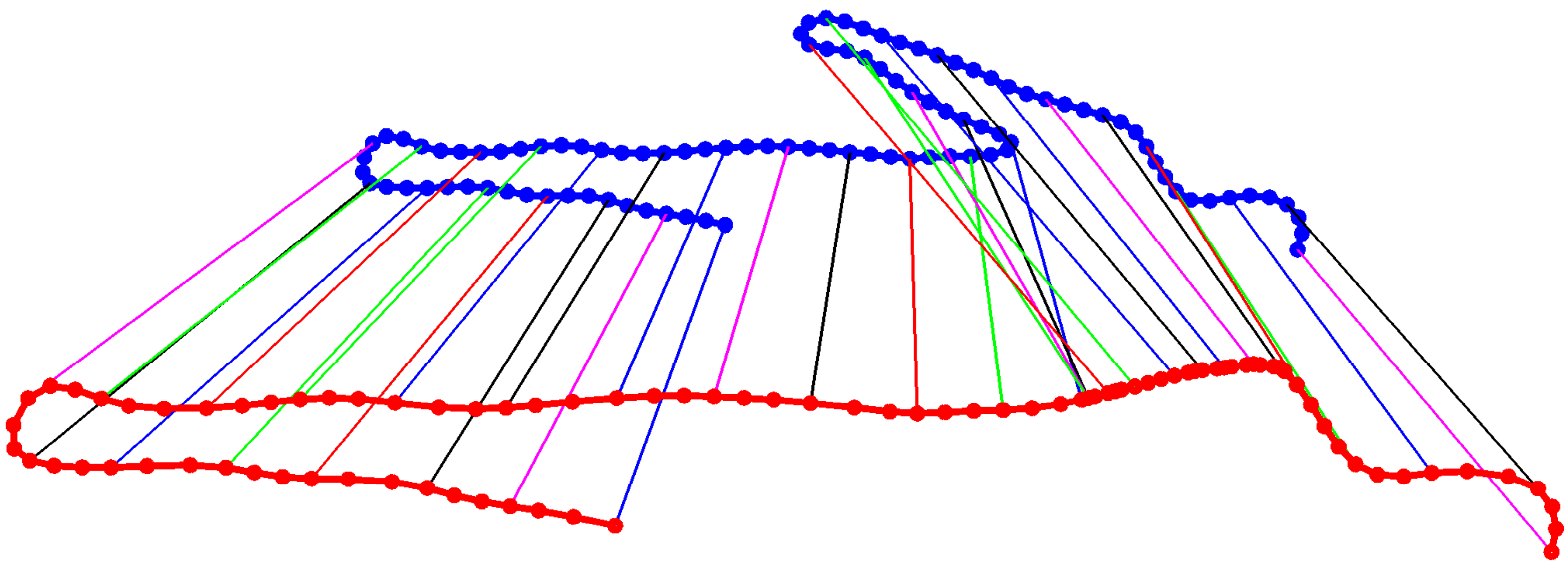}\\
			\small{(a) Fixed parameterization.} & \small{(b) Parameterization-invariant.}
		\end{tabular}
	\caption{\label{fig:rigid_vs_elastic}A toy example showing the importance of the invariance to re-parameterization.  (a) Correspondence under fixed parameterization. (b) Correspondence using a re-parameterization-invariant metric of~\cite{srivastava2011shape}. }
	}
\end{figure}

\subsubsection{Elastic registration as a reparameterization problem}

Instead of using landmarks, some previous works tried to find the optimal diffeomorphism $\diffeo^*$ that puts $f_2$ in full correspondence with $f_1$. Since the search space for optimal re-parameterizations is large,
some  of the previous works  such as  SPHARM~\cite{brechbuhler1995parametrization,kelemen1999elastic} and  SPHARM-PDM~\cite{gerig2001shape,styner2006framework}, used a fixed parameterization that is analogous to the arc-length parameterization on curves.  Others  restricted the type of re-parameterizations to a specific subset of the space of diffeomorphisms.  Lipman et al.~\cite{Lipman:2009:MVS}, for example,  showed that  the  optimal correspondence between nearly-isometric genus-0 surfaces, conformally embedded onto a sphere,  can be found by searching for the best M$\ddot{o}$bius transform that aligns them. The approach is, however, restricted to nearly isometric surfaces. In addition, it only finds a sparse set of correspondences since complex surfaces cannot be conformally embedded onto a sphere with minimal distortion. Zeng et al.~\cite{zeng_cvpr2010} extended Lipman et al.'s  approach and finds a dense matching in the presence of large elastic deformations.  Kim et al.~\cite{Kim:2011:BIM}  showed that by blending multiple maps, each one is a M$\ddot{o}$bius transform, one can match surfaces under relatively large elastic deformations. Nevertheless, the correspondences with these methods are not one-to-one. The inter-surface  maps they produce are not guaranteed to be a diffeomorphism. Thus they are not suitable for statistical shape analysis tasks.  In addition, these methods are limited to surfaces that can be conformally parameterized onto a sphere. Although, in theory, any closed genus-0 surface can be conformally embedded into a sphere, in practice, however, the distortion that results from conformal embedding of complex surfaces that have elongated extruding parts, \eg  human or animal body shapes, is often very large. As a consequence, several important shape parts are being under-sampled  and thus cannot be correctly matched across surfaces.


Using a fixed set of landmarks  or a fixed parameterization of the surfaces being analyzed imposes constraints on the correspondences that can be found and often may lead to undesirable results especially in the presence of large elastic deformations. 
Fig.~\ref{fig:rigid_vs_elastic}-(a) shows an example where  fixed parameterization fails. Better results can be  achieved by   treating the boundaries of objects as continuous surfaces, rather than discretizing them into point sets at the outset, and solving directly Eqn.~\eqref{eq:registration_problem} for finding the optimal  sampling (\ie reparameterization $\diffeo$) that better matches features across shapes as proposed in~\cite{kurtek2010novel,kurtek:mi2011,kurtekipmi2011,kurtek_mmbia2012,kurtek:pami2012,Jermyn:2012:ESM,Kurtek:CGF12063}, see Fig.~\ref{fig:rigid_vs_elastic}-(b) for a 2D example. Continuous  representations make it possible to elastically model and optimize correspondences.  The process of matching two 3D shapes  is, in fact, reduced to the problem of optimally re-parameterizing one shape so that it is as close as possible to the other.  Thus, correspondence becomes a re-parameterization problem, which can be mathematically formulated using diffeomorphisms. 

In order to do that, however, the action of $\diffeos$ on the space of surfaces should be by isometries, under the metric $d_{\preshapes}(\cdot, \cdot)$  \ie 
	\begin{equation}
		\forall \diffeo \in \diffeos, d_{\preshapes}(f_1\circ \gamma, f_2 \circ \gamma) = d_{\preshapes}(f_1, f_2).
		\label{eq:isometry_of_diffeos}
	\end{equation}
Kurtek et al.~\cite{kurtek2010novel,kurtek:mi2011,kurtekipmi2011,kurtek_mmbia2012,kurtek:pami2012,Jermyn:2012:ESM} showed that when using the $\ltwo$ metric in the space of surfaces $\preshapesf$, the condition of Eqn.~\eqref{eq:isometry_of_diffeos} is only valid when $\diffeo$ is an area-preserving diffeomorphism. This is, however, too restrictive especially when dealing with elastic registration because elastic deformations correspond to changes in surface area.

Kurtek et al.~\cite{kurtek2010novel} and Jermyn et al.~\cite{Jermyn:2012:ESM} showed that in the space of Square Root Maps (SRM) and in the space of Square-Root Normal Fields (SRNF), respectively, the action of $\Gamma$ is by isometries under the $\ltwo$ metric. Given that the $\ltwo$ metric in the space of SRNFs is equivalent to the partial elastic metric in the space of $\preshapesf$, one can 
\begin{itemize}
	\item Option 1: perform the analysis in $\preshapesf$ using the pullback of the partial elastic metric since the pullback metric inherits all the properties of the metric, or  
	\item Option 2: map the surfaces being analyzed to the space of SRNFs, perform the analysis there, \eg  registration and geodesic computation, and finally map, if needed, the results back to the space of surfaces $\preshapesf$.
\end{itemize}
In practice, option 2 is computationally very efficient; We are given two surfaces $f_1$ and $f_2$, and their SRNF representations $q_1= Q(f_1)$ and $q_2=Q(f_2)$. We  perform  elastic registration by solving for the optimal rotation $O$ and optimal re-parameterization $\gamma$ such that  the $\ltwo$ distance between $q_1$ and $O(q_2, \gamma)$ is minimized.  
If  $O^*$ and $\gamma^*$ is the solution to this optimization problem, then  $\bar{f}_2= O^*(f_2\circ \gamma^*)$ and its corresponding SRNF is given by $\bar{q}_2 = O^*(q_2, \gamma^*)$. 
Since this process uses the $\ltwo$ metric, geodesics correspond to straight lines in the space of SRNFs and thus one can use all the standard computational tools available when dealing with Euclidean spaces.  




\subsection{Geodesics}

\label{sec:geodesics}

For clarity purposes, let's assume for now that we are given two surfaces $f_1$ and $f_2 \in \preshapesf$ with a fixed parameterization. In other terms, a one-to-one correspondence between the surfaces is given and we are assuming  that this correspondence is correct.    The goal is to solve the inner optimization problem of Eqn.~\eqref{eq:geodesic_correspondence}, \ie  find a path $\path^*: [0, 1] \to \preshapesf$ that minimizes Eqn.~\eqref{eq:continuous_path_length}. Solutions to this optimization problem depend on the choice of the metric and  the representation (or nature of the pre-shape space).  When using the $\ltwo$ metric on $\preshapesf$, the solution is straightforward and in fact it can be expressed analytically as $
	F^*(\tau) = f_1 + \tau (f_2 - f_1), \text{with } \tau \in [0, 1].
$
As shown in Figure~\ref{fig:geod-syn}-(a), using the $\ltwo$ metric in $\preshapesf$ leads to undesirable distortions.  In this section, we will present the  main  strategies for computing geodesic paths when using elastic metrics that are physically motivated. We will also discuss details that depend on the choice of the metric and the representation.


\subsubsection{Geodesics using pullback metrics}  
\label{sec:geodesics_pullback}

The first class of approaches find geodesics by solving the optimization problem of Eqn.~\eqref{eq:geodesic_path_optimization} using one of the elastic metrics defined in Section~\ref{sec:elastic_metrics}. In the simplest case, we are given two surfaces $f_1$ and $f_2$ and we seek to find another surface $f^*$ on the geodesic path $F$ and which is exactly at the mid-point between $f_1$ anf $f_2$, \ie $f^* = F(0.5)$. Mathematically, $f^*$ is the minimizer of the discrete path energy:
\begin{equation}
	f^* = \arg\min_{f} L(F) = \energy(f_1, f) + \energy(f, f_2).
\end{equation}
which corresponds to solving for $f$ the Euler-Lagrange equation
\begin{equation}
	\nabla \energy(f_1, f, f_2) = \frac{\partial \energy(f_1, f)  }{ \partial f } + \frac{ \partial \energy(f, f_2)  }{ \partial f}  = 0.
\end{equation}
Heeren et al.~\cite{Heeren:2012:TGS} solve this non-linear optimization problem using the   Quasi-Newton method. That is, one can start with an initial estimate $f^0$  of $f$, \eg  $f^0 = 0.5 (f^1 + f^2)$, and then iteratively update the estimate via a Newton iteration with step size control:
$
	f^{n+1} = f^{n} + \epsilon \Delta f
$ 
 where
$$
	\Delta f = -\left[  \Hess \energy(f_1, f^n, f_1)   \right]^{-1} \nabla \energy.
$$
The optimal step size $\epsilon$ can be determined using the David-Fletcher-Powell (DFP) formula~\cite{davidon1991variable}. Using this formulation, one can  solve for the entire time-discrete geodesic path recursively; once the mid-point $f$ between $f_1$ and $f_2$ is computed, one can apply the same procedure to solve for the mid-point between $f_1$ and $f$, and between $f$ and $f_2$, etc.~\cite{Heeren:2012:TGS}. Alternatively, one can directly find the entire discrete path $F = \{ f_1= f^1, f^2, f^3, \cdots, f^K= f_2\}$, by optimizing
$$
\energy(f^i, i= 1, \cdots, K) = \sum_{i=1}^{K-1}  \energy(f^i, f^{i+1}),
$$
over  $f^i, i= 2, \cdots, K-1$. This is called a \emph{path straightening} procedure. The optimization can start with an initial path, which is then iteratively straightened, \ie updated so that its length with respect to the elastic metric is minimized.    In the space of shells~\cite{Heeren:2012:TGS,zhang2015shell}  this is done by   solving the set of Euler-Lagrange equations:
$$
	\frac{ \partial \energy }{ \partial f^{i} } = 0, i=2, \cdots, K-1.
$$
Instead of solving a non-linear optimization problem, Kurtek et al.~\cite{kurtek:pami2012,samir2014elastic}  define the energy of a continuous path $F$ as:
\begin{equation}
	E_{path}(F) = \int_0^1 \innerd{  \frac{ d F}{d\tau}(\tau)   }{\frac{d F}{d\tau}(\tau)  }   d\tau.
	\label{eq:path_energy} 
\end{equation} 
The shortest path is then found by minimizing the path energy using a gradient descent approach where the current estimate of the geodesic path is updated according to $F = F - \epsilon \nabla E_{path}$. Let $\mathcal{A}_0$ be the set of all smooth paths that start at $ f_1$ and end at $f_2$ and let $T_F(\mathcal{A}_0)$ be the tangent space to   $\mathcal{A}_0$ at a given point $F$.  $T_F(\mathcal{A}_0)$ corresponds to perturbations of the path $F$. 
Kurtek at al.~\cite{kurtek:pami2012} describes the space of path perturbations using orthonormal basis $\mathcal{B} = \{ \alpha_1, \cdots, \alpha_B \}$ so that the gradient of the path energy can be written as
$$
\nabla E_{path} = \sum_i \nabla E_{path} (\delta \alpha_i) \delta \alpha_i.
$$ 
We refer the reader to~\cite{kurtek:pami2012,samir2014elastic} for the full derivation of the directional derivatives $\nabla E_{path} (\delta \alpha_i) $ as well as the procedure for constructing the orthonormal basis $\mathcal{B}$.

\subsubsection{Geodesics in the space of SRNFs}
\label{sec:srnf_geodesics}
 
The SRNF representation, compared to others,  is very promising as a representation of surfaces for elastic shape analysis. If geodesics, mean shapes, PCA, etc.\ could be computed in $\preshapesq$ and $\shapesq$ under the $\ltwo$ metric, and then mapped back to $\preshapesf$,  there would be large gains in computational efficiency with respect to the optimization-based methods described in the previous sections and those introduced in \eg~\cite{Jermyn:2012:ESM,xie-iccv:2013,zhang2015shell}.  Unfortunately,  unlike the curve case, there is no analytical expression for $Q^{-1}$, the inverse of SRNF map $Q$, for arbitrary points in ${\preshapesq}$. Moreover, the injectivity and surjectivity of $Q$ remain to be determined, meaning that for a given $q\in\srnfs$, there may be no $f\in\preshapesf$ such that $Q(f) = q$, and if such an $f$ does exist, it may not be unique. 
If one cannot invert the representation, one can transfer geodesics and statistical analyses conducted in $\preshapesq$ back to $\preshapesf$ by pulling the $\ltwo$ metric back to  $\preshapesf$ under and perform computations there, as in~\cite{xie-iccv:2013}, but this is computationally expensive, and rather defeats the purpose of having an $\ltwo$ metric in the first place.


To solve this inversion problem, Xie et al.~\cite{xie2014numerical} showed that in the case of star-shaped genus surfaces, an analytical solution to the inversion problem exists. Laga et al.~\cite{laga2016numerical}, on the other hand,    developed a method that, given $q\in\preshapesq$, finds an $f\in\preshapesf$ such that $Q(f) = q$, if one exists, or an $f$ whose image $Q(f)$ is the closest, in terms of the elastic metric, to $q$ if it does not. 
This is achieved by formulating SRNF inversion as an optimization problem: find an element $f \in \preshapesf$ whose image $Q(f)$ is as close as possible to the given $q\in \preshapesq$ under the $\ltwo$ norm.  This optimization problem is efficiently solved using a gradient descent approach~\cite{laga2016numerical}. To avoid undesirable solutions, since gradient descent procedures converge to local minima if not initialized appropriately, Laga et al.~\cite{laga2016numerical} 
 carefully engineered an orthonormal basis of $\preshapesf$, and used a spherical-wavelet based multiresolution and multiscale representation of the elements of $\preshapesf$. The basis can be generic, \eg  spherical harmonics, or domain-specific such as PCA basis in the presence of training datasets.  With this approach,  SRNF maps can be inverted with high accuracy and efficiency, with robustness to local minima, and with low computational complexity.  

Finally, the shapes on the geodesic between two star-shaped surfaces are not necessarily star-shaped, and thus,  the analytic form may not be the correct inversion. However, one can use the analytical solution   as an initial guess for the inverse, thereby better initializing the reconstruction-by-optimization problem.

By mapping back to $\preshapesf$ geodesics computed in $\shapesq$, which are straight lines since they are equipped with the $\ltwo$ metric, the  computational cost is reduced by  an order of magnitude compared to methods that use path straightening under the pullback metric.  
See Section~\ref{sec:examples_applications}  for a few examples of geodesics between pairs of complex genus-0 surfaces.



\begin{figure}[!t]
    \centering
    		 \includegraphics[width=.48\textwidth]{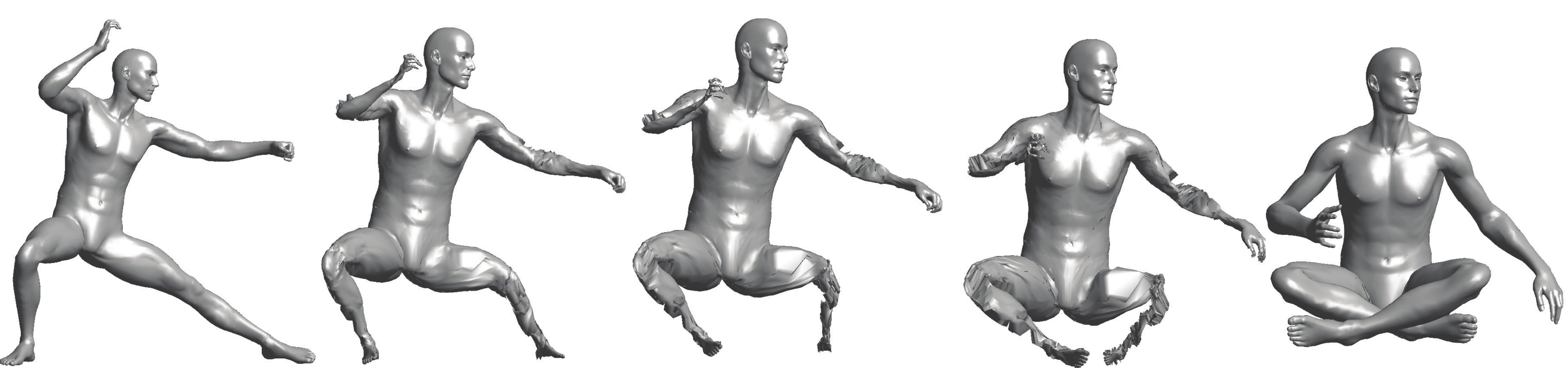}\\
		 \small{(a) Geodesic with the $\ltwo$ metric, i.e. linear path $(1-t)f_1 + t f_2$. The registration computed with functional maps.}
		  \includegraphics[width=.48\textwidth]{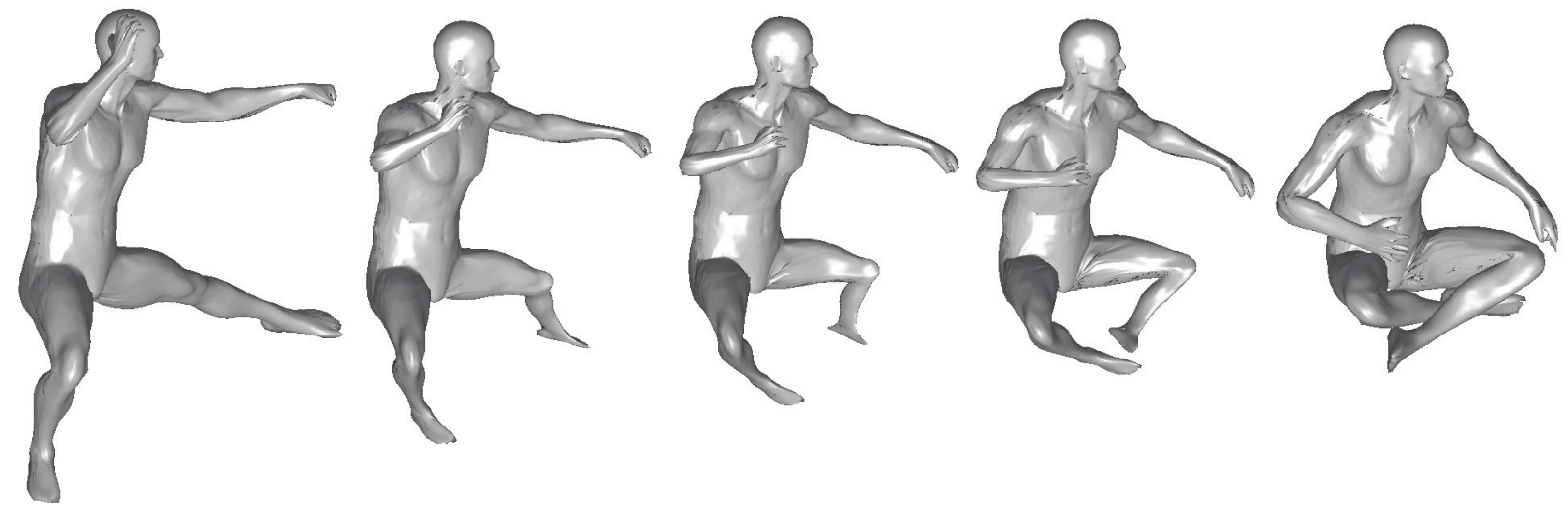}\\
		  \small{(b) Geodesic with the $\ltwo$ metric, i.e. linear path $(1-t)f_1 + tf_2$. The registration computed with SRNF.}\\ 
		  \includegraphics[width=.48\textwidth]{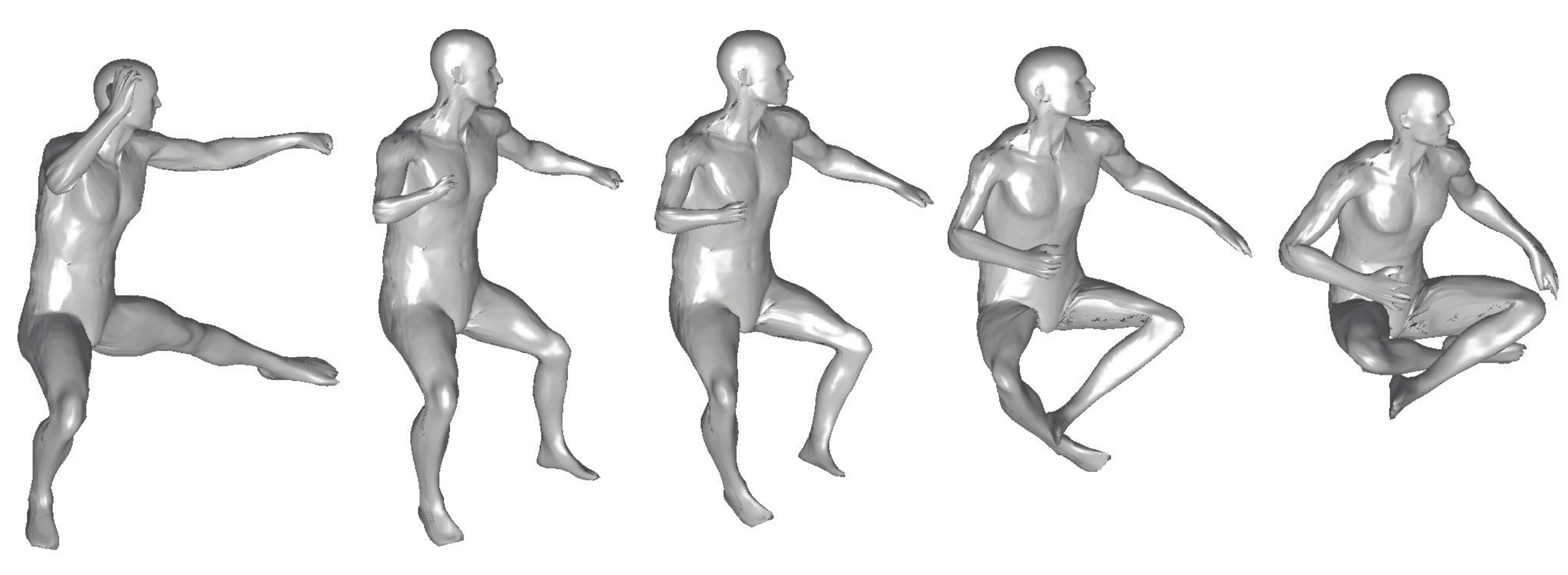}\\
		  \small{(c) Geodesic path using SRNF inversion.} \\
		 \vspace{6pt}
		 \includegraphics[width=.15\textwidth]{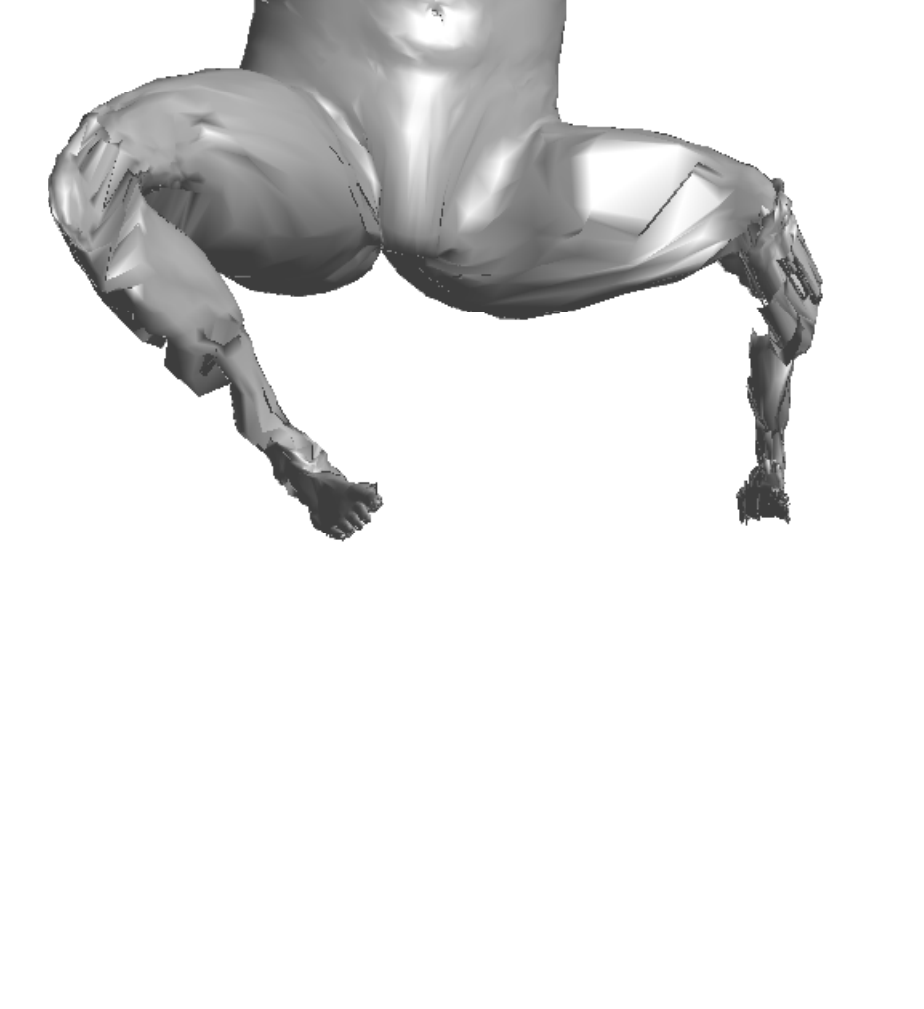}
		 \includegraphics[width=.15\textwidth]{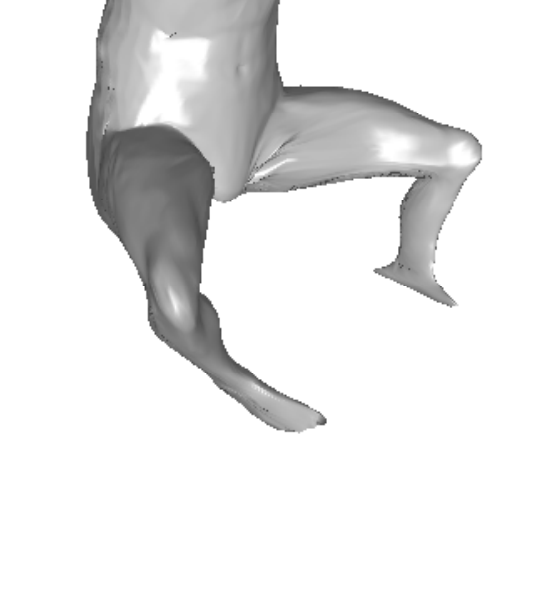} 
		 \includegraphics[width=.15\textwidth]{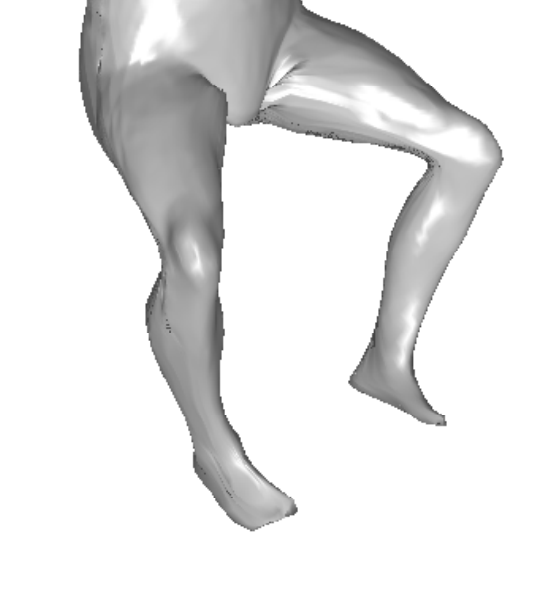}\\
		 \vspace{-8pt}
		 \small{Zoom on (a)} \hspace{1cm}  \small{Zoom on (b)} \hspace{1cm}  \small{Zoom on (c)}  \\

    \caption{\label{fig:geodesics_example1} Comparison between correspondence and geodesic computation algorithms. In (a) correspondences are computed using functional maps~\cite{Ovsjanikov:2012:FM}, while the deformation is computed using linear interpolation in the preshape space $\preshapesf$. In (b), the registration is computed using the re-parameterization invariant elastic metric in the space of SRNFs while the deformation is computed using linear interpolation in he preshape space $\preshapesf$. Finally, in (c), the registration is computed using the re-parameterization invariant elastic metric in the space of SRNFs while the geodesic is  obtained by SRNF inversion of linear paths in $\shapesq$.  See~\cite{laga2016numerical} for more details.}
\end{figure}

\begin{figure}[!t]
    \centering
		  \includegraphics[width=.48\textwidth]{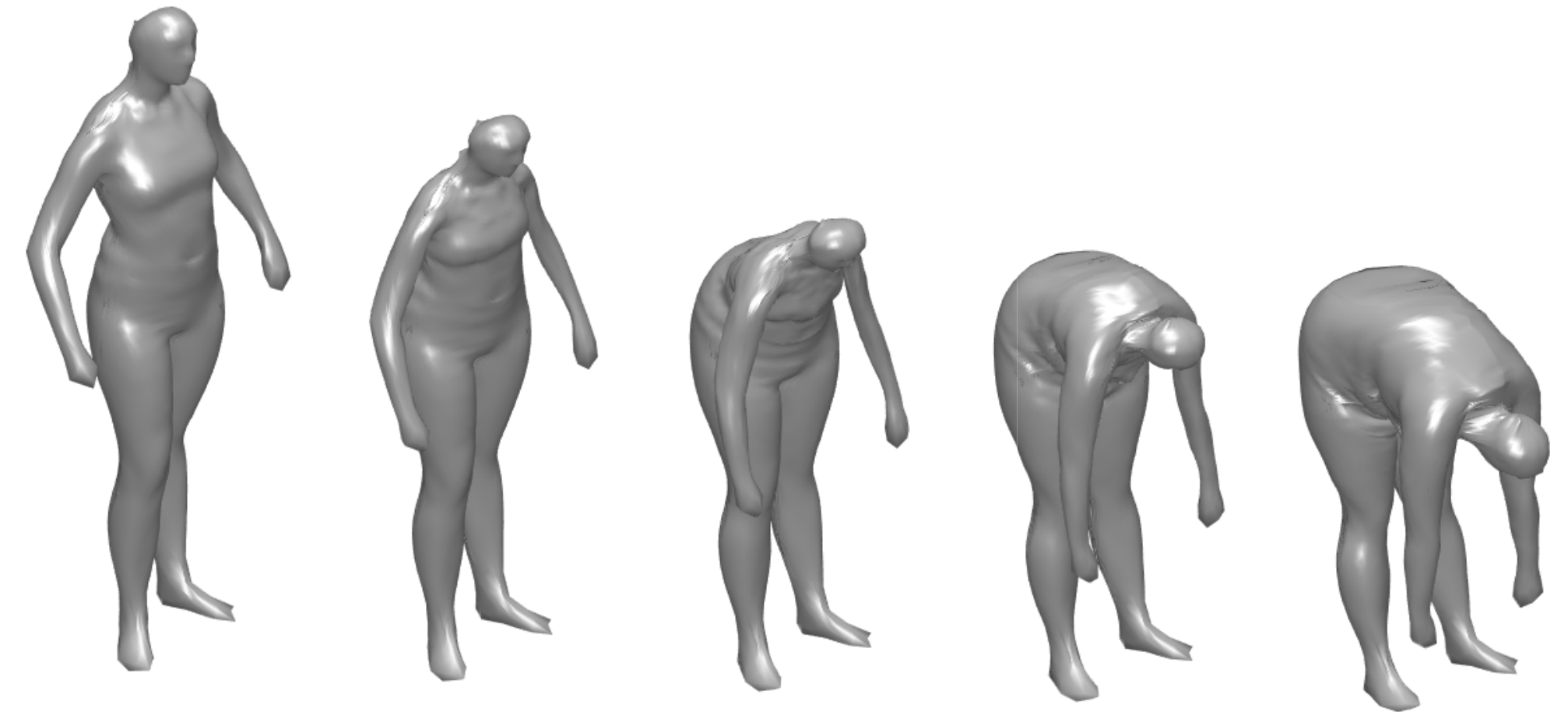} \\
		  \small{(a) Linear path $(1-t)f_1 + tf_2$, registration computed with SRNF.}\\
		  \includegraphics[width=.48\textwidth]{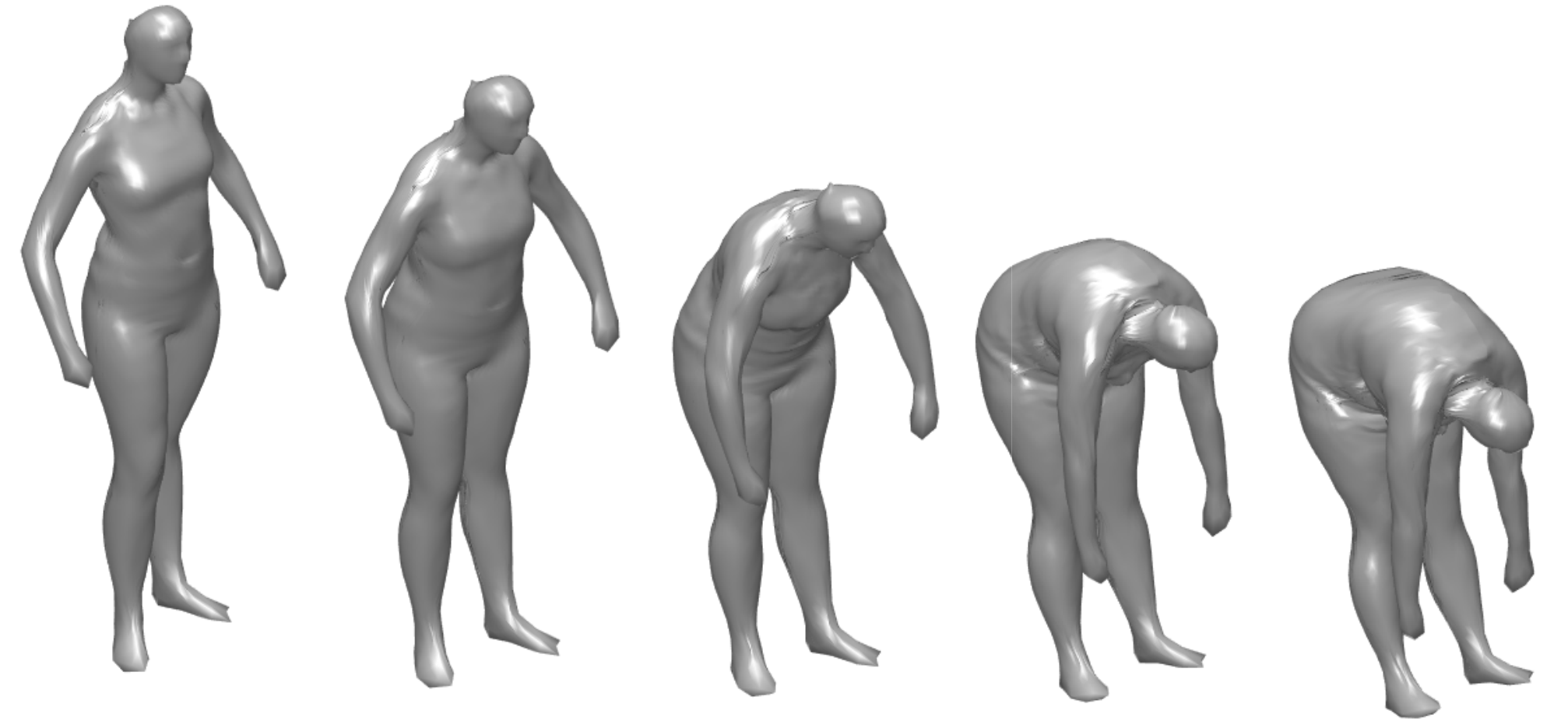}\\
		  \small{(b) Geodesic path using SRNF inversion.}		
    \caption{\label{fig:geodesics_example2} Comparison between (a)  linear paths in $\shapesf$, and (b) geodesics obtained by SRNF inversion of linear paths in $\shapesq$.   See~\cite{laga2016numerical} for more details. }
\end{figure}


\subsubsection{Comparison and discussion}
As the reader can see, many shape spaces and metrics have been defined in the literature. To help the reader navigate in this rich literature, we summarize in this section their properties and discuss their similarities and differences.

The first  aspect to consider when choosing a metric and a shape space is  the invariance properties of the metric, which defines the type of pre-processing requirements. For  instance, as summarized in Table~\ref{tab:taxonomy_equations},  the metrics of Kilian et al.~\cite{Kilian:2007:GMS} as well as those defined in the space of thin-shells~\cite{Heeren:2012:TGS,heeren2016splines,BeFlHe13,zhang2015shell}  are invariant to translation, scale, and rotation and thus one does not need to normalize the input 3D models in a pre-processing step. On the other hand, diffeomorphisms do not act by isometries in these shape spaces under the defined metrics. As a consequence, these metrics are not invariant to re-parameterization and thus are not suitable for joint registration and geodesic computation.  On the other hand, the $\ltwo$ metric on the space of square-root maps such as the SRNF~\cite{Jermyn:2012:ESM} and Q-maps~\cite{kurtek:pami2012} are invariant to re-parameterization. Thus the correspondence and registration problem can be solved jointly, using the same metric, with the geodesic computation. The SRNF requires normalization for scale while the Q-maps require normalization for translation and scale. Finally, methods based on M-reps~\cite{fletcher2004principal} and Lie bodies~\cite{Freifeld:2012:LBM} require normalization for all rigid transformations as well as the pre-computation of correspondences using a different approach.

The other aspect to consider is the types of deformations that are quantified, and subsequently,  the complexity of the metric and the representation, which affect the processing time. Ideally, the process of evaluating the metric should be fast so that geometry processing and shape analysis tasks can be performed in a computationally reasonable time. 
From this aspect, Euclidean shape spaces and Euclidean metrics are probably the most convenient to use, since evaluating the $\ltwo$ metric, computing short paths, and performing statistics under the metric are straightforward. Moreover, when working in Euclidean spaces one can  use all the computational tools developed for vector spaces, \eg      interpolations and extrapolations, Principal Component Analysis (PCA), and probabilistic models. As such, the SRNF representation~\cite{Jermyn:2012:ESM} is very convenenint since all these tasks can be performed in the SRNF space using the $\ltwo$ metric, which is equivalent to the partial elastic metric in $\preshapesf$, and map the results back to $\preshapesf$ for visualization~\cite{xie2014numerical,laga2016numerical}.

\section{Statistical analysis under elastic metrics}
\label{sec:statistical_analysis}

At this stage we have all the necessary computational ingredients, which are a shape  space, a metric,  and a mechanism for computing  correspondences and geodesics. In this section, we will show how to use these ingredients for performing full statistical  analysis of the shape of 3D objects, \ie given an input collection of 3D models, we want to
\begin{itemize}
	\item compute the mean shape and the modes of variations. This is often referred to as the \emph{the shape atlas},  
	\item characterize a population of 3D shapes with a probability distribution, and finally 
	\item synthesize arbitrary shapes by sampling from the fitted probability distribution. 
\end{itemize}
We have already seen in Section~\ref{sec:morphable_models} that when the metric is Euclidean, in the case of morphable models for example, computing such quantities is straightforward using standard tools in vector spaces. In this section, we focus on the case where the shape space and the metric are non-linear. 

\subsection{Statistical analysis using non-Euclidean metrics}

When using non-Euclidean metrics,  Eqn.~\eqref{eq:karcher_mean_shapespace}, which is equivalent to Eqn.~\eqref{eq:karcher_mean_registered}, cannot be solved analytically. Several previous works have proposed 
gradient-based approaches  for finding the Karcher mean of Eqn.~\eqref{eq:karcher_mean_registered}, see for example~\cite{dryden-mardia_book:98,le-mean-shape}. For convenience, the general approach is  repeated here as  Algorithm~\ref{alg:karcher_mean_registration}.

\begin{algorithm}[!ht]

 \caption{\label{alg:karcher_mean_registration}Joint Karcher mean computation and registration.}
 
 \textit{Input:} Set of surfaces $\{f_1,\dots,f_n\}\in{\surfaces}$, step size $\epsilon_1$, and energy difference threshold $\epsilon_2$.\newline
 \textit {Output:} Karcher mean surface $\meansurf$. 

\begin{algorithmic}[1]

 \STATE For each $i=1,\dots,n$, normalize the translation and scale of $f_i$. Let's denote by $f_i$  the normalized surfaces. 
 \STATE Set $j=0$ and initialize the mean shape as $\meansurf_j  =  f_0$.
 
  \STATE \label{step:registration} For each $i=1,\dots,n$, register $f_i$ to $\meansurf_j$ resulting in $f^*_i=\rotation^*(f_i\circ\diffeo^*)$, where $\rotation^*$ and $\diffeo^*$ are the optimal rotation and reparameterization, respectively.
 \STATE For each $i=1,\dots,n$, compute the shape space geodesic path $\geodf_i$ between $\meansurf_j$ and $f^*_i$.
 \STATE Compute the average shooting direction as $v=(1/n)\sum_{i=1}^n \deriv{\geodf_i}{t}|_{t=0}$.
 \STATE If $\norm{v}<\epsilon_2$, stop and return $\meansurf_j$. Otherwise, update the mean using $\meansurf_{j+1}=\meansurf_{j}+\epsilon_1 v$ and normalize $\meansurf_{j+1}$ with respect to scale and translation.
 \STATE Return to step 3.
 
\end{algorithmic}
 
\label{alg:joint_karcher_mean_registration}
\end{algorithm}

In addition to computing the mean shape of a collection of 3D models, we are usually interested in understanding (and modelling) how the shape varies within the collection.  Let $\{f_1, \cdots, f_n\}$ be a collection of shapes and $\meansurf$ be their mean. For simplicity of presentation, we assume that all the surfaces have been registered to $\meansurf$. Let $\alpha_i$ be the geodesic path, with respect to the chosen metric, between $\meansurf$ and $f_i$. Fletcher et al.~\cite{fletcher2004principal} defined the following concepts that are essential for representing variability of data that lie on general manifolds:
\begin{itemize}
	\item Variance. Flechet~\cite{frechet1948elements} defined variance as  the expected value of the squared Riemannian distance from the mean.
	\begin{equation}
		\sigma^2 = \frac{1}{n}\sum_{i=1}^n d_{\shapes} ([\meansurf], [f_i])^2. 
		\label{eq:non_linear_modes}
	\end{equation}	
	\item Geodesic subspaces. The lower-dimensional subspaces in PCA are linear subspaces. For general manifolds, it is natural to use geodesic curves as one-dimensional subspace, which is equivalent to principal directions in PCA.  Flechet~\cite{frechet1948elements} showed that  submanifold geodesics at the mean  are the generalization of the linear subspaces of PCA.	
	\item Projection. The projection of a point $f \in \preshapesf$ onto a geodesic submanifold $H$ is defined as the point on $H$ that is the nearest to $f$ in terms of  the Riemannian distance. Thus, the projection
operator $\pi_H(f): \preshapesf \to H$  can be defined as:
	\begin{equation}
		  \pi_H(f) = \argmin_{[y] \in H} d_{\shapes}([f], [y] )^2.	
		  \label{eq:projection_operator}	
	\end{equation}
\end{itemize}
Given the non-linearity of the shape space, the modes of variation can be efficiently computed by first projecting the data to the tangent space to $\preshapesf$ at the Karcher mean using the $\exp$ map, performing standard PCA in that linear space, and then mapping the results back to $\preshapesf$ using the Log map. Algorithm~\ref{alg:non_linear_pca_statistics} summarizes this procedure. 
 
\begin{algorithm}[!ht]

 \caption{\label{alg:non_linear_pca_statistics}Modes of variation and random shape sampling.}
 
 \textit{Input:} Set of surfaces $\{f_1,\dots,f_n\}\in{\preshapesf}$ normalized and registered to their Karcher mean $\meansurf$.  \newline
 \textit {Output:} Modes of variation and random shapes. 

\begin{algorithmic}[1]
	\STATE Project the geodesic paths $\geodf_i$, which connect each surface $f_i$ to $\meansurf$,  onto  $T_{\meansurf}(\preshapesf)$, the tangent space to $\preshapesf$ at $\meansurf$. This is done using the Log map. Let $v_i = \text{Log}(\alpha_i)$.
	\STATE Compute the covariance matrix $K = \frac{1}{n-1} \sum_i v_i v_i^T$.
	\STATE Compute the standard eigenvalues and eigenvectors of $K$.
	\STATE Any tangent vector $v$ in $T_{\meansurf}(\preshapesf)$ can be written as $v = \sum_{i=1}^k \lambda_i V_i$, where $k\le n$ is the number of leading eigenvectors of $K$ and $V_i$ is the $i-$th leading eigenvector. $\lambda_i$ are real coefficients.
	\item A new shape $f$ is then obtained by projecting $v$  onto $\preshapesf$ using exponential map.
 
\end{algorithmic}

\end{algorithm}
Note that these two algorithms for computing the Karcher mean and the modes of variation are generic and can be used with any representation of surfaces and metrics. They both involve co-registration, geodesics computation, and the computation of the exponential and log maps.  The former can be omitted when the surfaces are pre-registered using a different procedure and different optimality criteria. This is however suboptimal.  The computation of geodesics, exponential maps and log maps depend on the metric and the shape space.  When using M-reps or Lie bodies of Section~\ref{sec:transformation-based representations},  geodesics and geodesic distances have analytical solutions.  In other cases, \eg when using the metrics in the space of shells or when using the square-root representations, one has to use the optimization procedures described in Section~\ref{sec:geodesics}.


\subsection{Statistical analysis by SRNF inversion}

\label{sec:statistics_srnf_inversion}

Computing shape summaries on non-linear shape spaces using complex elastic metrics, or deformation energies, requires computing geodesics many times. For instance, at each iteration of  Algorithm~\ref{alg:karcher_mean_registration}, $n$ geodesics should be computed where  $n$ is the number of shapes in the collection to average. This is computationally very expensive as discussed in~\cite{laga2016numerical}. SRNF representations offer an elegant and a computationally efficient alternative solution; 
Instead of using the pullback metric, one can simplify the iterative procedure given in Algorithm \ref{alg:karcher_mean_registration} by first mapping all surfaces to the SRNF space, $\preshapesq$, resulting in $\{\srnf_1,\dots,\srnf_n\}$. Then, the mean shape in SRNF space, denoted by $\bar{\srnf}$, is computed by iterating between co-registering all $\srnf_i$s to $\bar{\srnf}$, and subsequently updating $\bar{\srnf}$. This is computationally very efficient since $\preshapesq$ is an Euclidean space. Finally, at the last step, the mean shape $\meansurf\in[\meansurf]$ is computed by SRNF inversion, \ie finding a shape $\meansurf$ such that  $\srnfmap(\bar{f}) = \bar{\srnf}$. The entire procedure is summarized in Algorithm~\ref{alg:karcher_mean_srnf_inversion} and detailed in~\cite{laga2016numerical}.

\begin{algorithm}

\textit{Input:} Set of surfaces $\{f_1,\dots,f_n\}\in{\preshapesf}$ and their SNRFs $\{\srnf_1,\dots,\srnf_n\}\in \preshapesq$.\newline
 \textit {Output:} Karcher mean surface $\meansurf$.
 
 \begin{algorithmic}[1]
\STATE Let $\bar{\srnf} = Q(\meansurf)$ with $\meansurf$ set to  $f_1$ as the initial estimate of the Karcher mean. Set $j=0$.

\STATE For each $i=1,\dots,n$, register $\srnf_i$ to $\bar{\srnf}$ resulting in $\srnf^*_i=\rotation_i^*(\srnf_i, \diffeo_i^*)$, where $\rotation_i^*$ and $\diffeo_i^*$ are the optimal rotation and reparameterization, respectively. 


\STATE Update the average $\bar{\srnf} = \frac{1}{n}\sum_{i=1}^n \srnf_i$.

\STATE If change in $\left\|\bar{\srnf}\right\|$ is large go to Step 2.

\STATE Find $\meansurf$ by SRNF inversion s.t. $Q(\meansurf) = \bar{\srnf}$. 

\STATE Return $\meansurf$, $\rotation_i$ and $\diffeo_i, i=1, \dots, n$. 

\end{algorithmic}

\caption{\label{alg:karcher_mean_srnf_inversion}Sample Karcher mean by SRNF inversion.}
\end{algorithm}

Next, let $\bar{\srnf}$ denote the SRNF of the final average shape. With a slight abuse of notation, let $\srnf_i,\ i=1,\dots,n$ denote the SRNFs of the surfaces in the sample optimally registered to this average. Then,  since the SRNF space is Euclidean, the modes of variation can be computed in a standard way, \ie using singular value decomposition (SVD) of the covariance matrix. Let $u_q^k$ denote the $k-$th principal direction. Then, one can explore the variability in this direction around the mean using $\bar{\srnf} + \lambda u_q^k$, where $\lambda \in \real$. To visualize the principal directions of variation in $\shapesf$, we only need to compute, by SRNF inversion, $f^k_\lambda$ such that $\srnfmap(f^k_{\lambda}) =\bar{\srnf} + \lambda u_q^k$.

Finally, given a sample of observed surfaces $f_1,\dots,f_n \in \preshapesf$, one wants to fit a probability model to the data and sample from it new 3D shape instances. Estimating probability models on nonlinear spaces like $\shapesf$ is difficult: classical statistical approaches, developed for vector spaces, do not apply directly. By first computing the SRNF representations of the surfaces, one can use all the usual statistical tools in the vector space $\shapesq$, and then map the results back to $\preshapesf$ using the  SRNF inversion algorithm.  Let $G(q)$ be the model probability distribution fitted to $\{q_1,\dots,q_n\}$.  A random sample $q_s$ can be generated from $G$. We then find $f_s$ such that $Q(f_s) = q_s$.  Note that $G$ can be an arbitrary distribution, parametric or non-parametric. Here, we use a wrapped truncated normal distribution, which can be learned using Principal Component Analysis on $\shapesq$.   We caution the reader that the distribution is defined and analyzed in SRNF space; a distribution on $\shapesq$ induces a distribution on $\preshapesf$, but it  has not been derived explicitly.

In terms of computational complexity, computing the mean shape only requires the inversion of one  SRNF, while   techniques based on the pullback metric such as~ \cite{kurtek2010novel,kurtek:pami2012,xie-iccv:2013} are iterative, and require the computation of $n$ geodesics per iteration, with $n$ being the number of shapes to average. This is computationally very expensive especially since each geodesic is computed using the expensive pullback metric.

\section{Examples and applications}
\label{sec:examples_applications}

We consider in this section a few examples of elastic 3D shape analysis tasks using the tools discussed in this chapter. Note that this section is not intended to provide a thorough performance evaluation and comparison of the methods described in this chapter but instead it aims to show their potential and the range of shape analysis problems that can be efficiently solved with these methods. 

\subsection{Registration and geodesic deformations}
The first example we consider is the joint  registration and geodesic deformation of 3D objects that undergo complex elastic deformations. Here, we would like to emphasize for the reader that elastic (or non-rigid) deformation is composed of bending and stretching. Works based on (generalized) multi-dimensional scaling~\cite{bronstein2006generalized} and (heat-)diffusion geometry~\cite{bronstein2008numerical,bronstein2010scale} only consider bending, which preserves intrinsic properties of surfaces. Here, we consider full elastic deformations. 

In Figure~\ref{fig:geodesics_example1}, we consider  two human body shapes of the same subject but in significantly different poses and compute a joint registration and geodesic deformation. We first compare the quality of the registration when computed with a re-parameterization invariant elastic metric (Figure~\ref{fig:geodesics_example1}-(b) and Figure~\ref{fig:geodesics_example1}-(c)) and when computed with other methods, \eg  functional maps of~\cite{Ovsjanikov:2012:FM}. As one can see in Figure~\ref{fig:geodesics_example1}-(a), the later method does not produce correct one-to-one correspondences compared to the methods that formulate registration as a re-parameterization problem. Next, we compare geodesics computed using the $\ltwo$ metric in $\preshapesf$ vs. geodesics computed using elastic metrics by SRNF inversion. This is shown in Figures~\ref{fig:geodesics_example1}-(b), (c) and Figure~\ref{fig:geodesics_example2}-(a), (b), where shapes undergo complex bending and stretching.

\begin{figure}[t]
    \centering

    	\includegraphics[width=.5\textwidth]{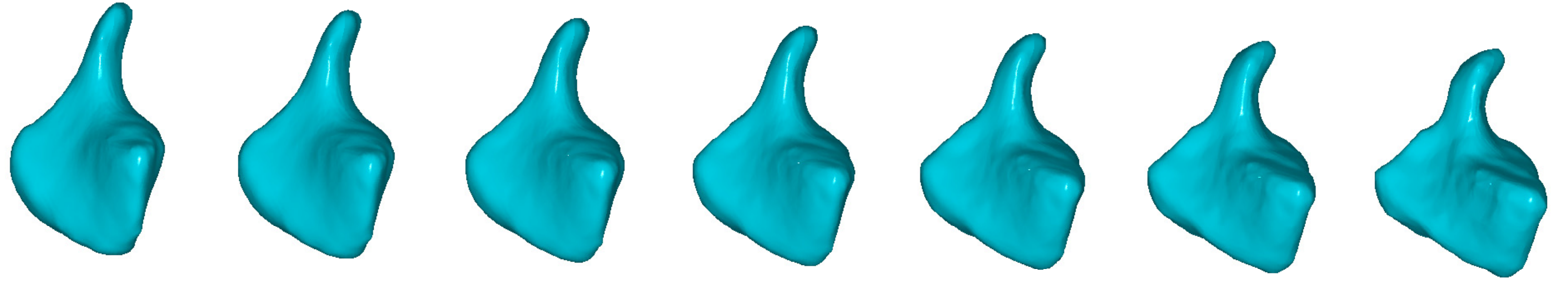} \\
	\includegraphics[width=.35\textwidth]{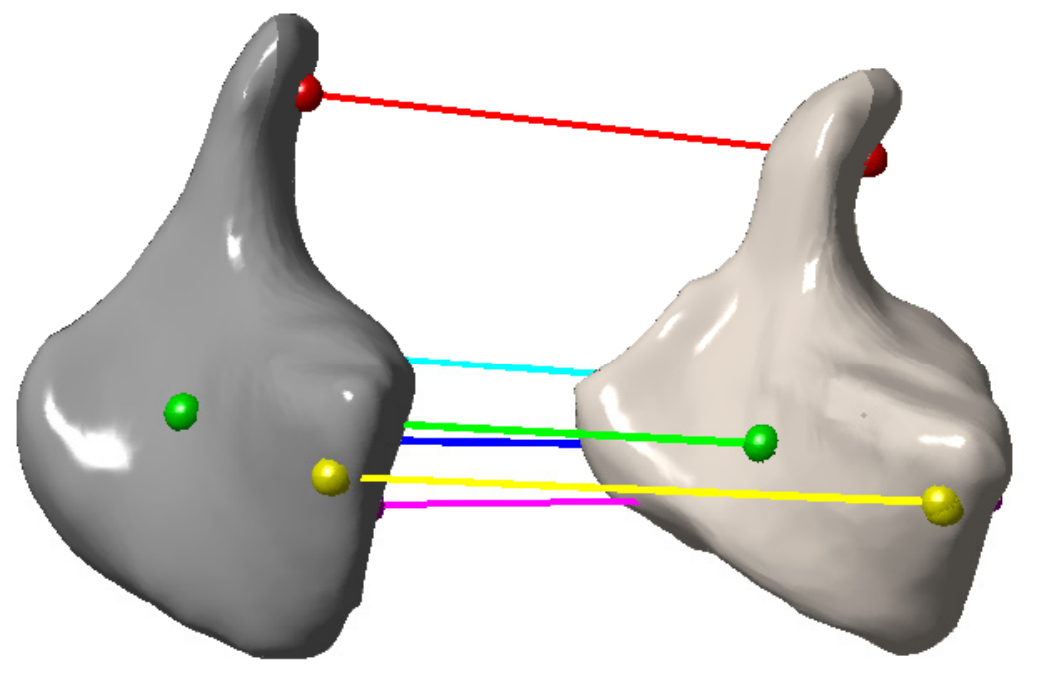} \\
	
    \caption{\label{fig:geodesic_bones} Elastic correspondence and geodesic deformation between two carpal bones (left), and illustration of a few landmark correspondences (right). See~\cite{banerjee2015generation} for more details. }
\end{figure}

\subsection{Elastic co-registration of 3D shapes}

\begin{figure}[t]
    \centering

    	\includegraphics[width=.48\textwidth]{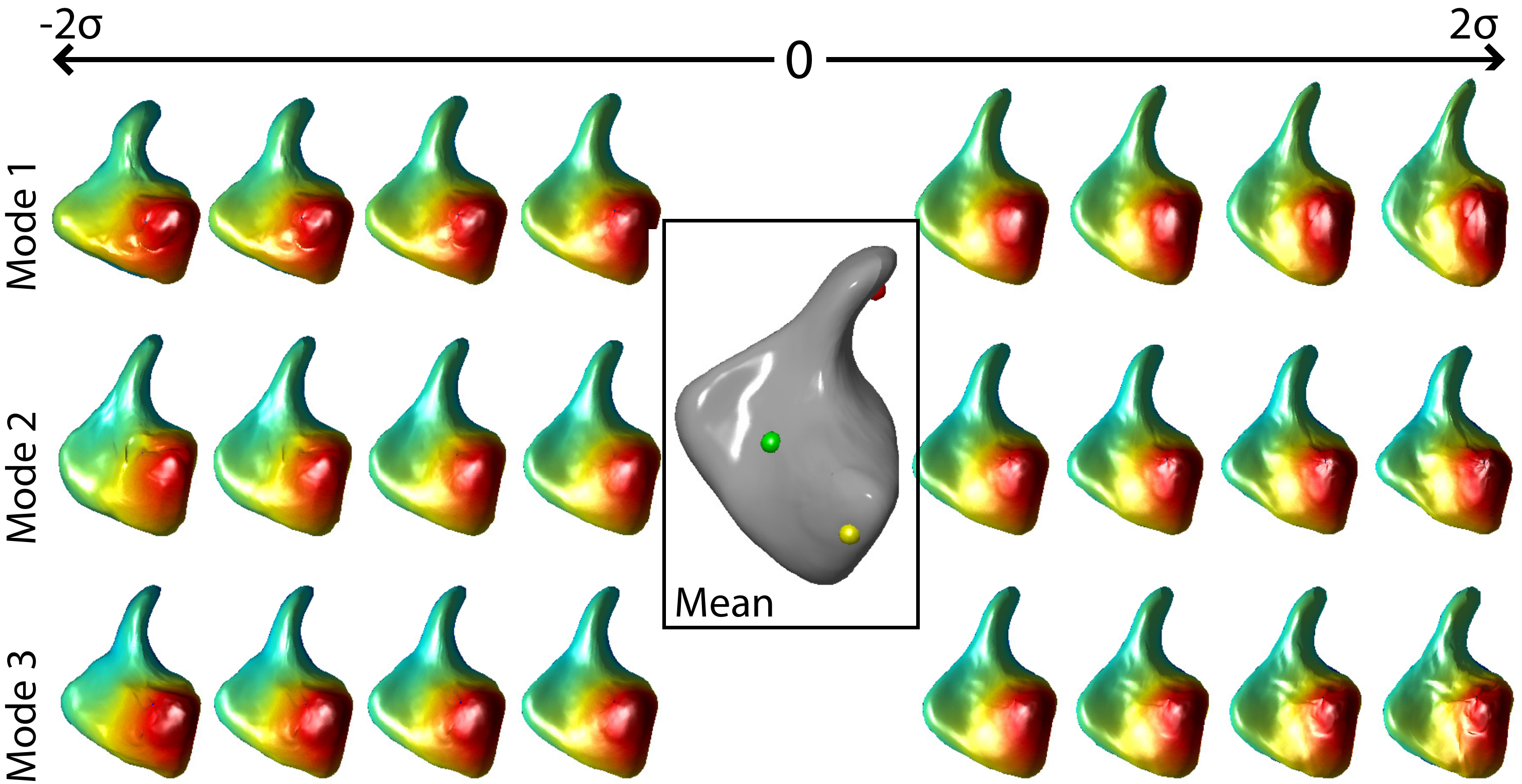} \\
	
    \caption{\label{fig:hamate_stats} Mean and three leading modes of variations of the Hamate carpal bone shape. Elastic correspondences and statistics computed using the SRNF and SRNF inversion framework. See~\cite{banerjee2015generation} for more details. }
\end{figure}

Standard registration methods are often pairwise, \ie they put in correspondence a pair of 3D objects. Here, in statistical shape analysis, we are dealing with multiple objects that should be registered to each other in a globally optimal way. Moreover, as discussed in Sections~\ref{sec:registration_geodeiscs} and~\ref{sec:statistical_analysis}, using a different criteria for registration, geodesics, and mean computation leads to suboptimal solutions.  An important by-product of the Karcher mean computation using a re-parameterization invariant metric is the simultaneous co-registration of multiple objects.  This is illustrated with the examples of Figures~\ref{fig:mean_modes_shape} and~\ref{fig:hamate_stats}. In the first example, we take multiple 3D human body models in different poses and belonging to different subjects, put them in correspondence while simultaneously computing their Karcher mean as described in Section~\ref{sec:statistics_srnf_inversion}.  We perform the same experiment on the example of Figure~\ref{fig:hamate_stats} this time taking 3D models of the hamate bone in the hand wrest, compute the mean shape while simultaneously registering all the 3D models in the corpus.

%

\subsection{Classification}
\begin{figure}[!t]
    \centering
    	\begin{tabular}{@{}c@{ }c@{}}
		  \includegraphics[width=.24\textwidth]{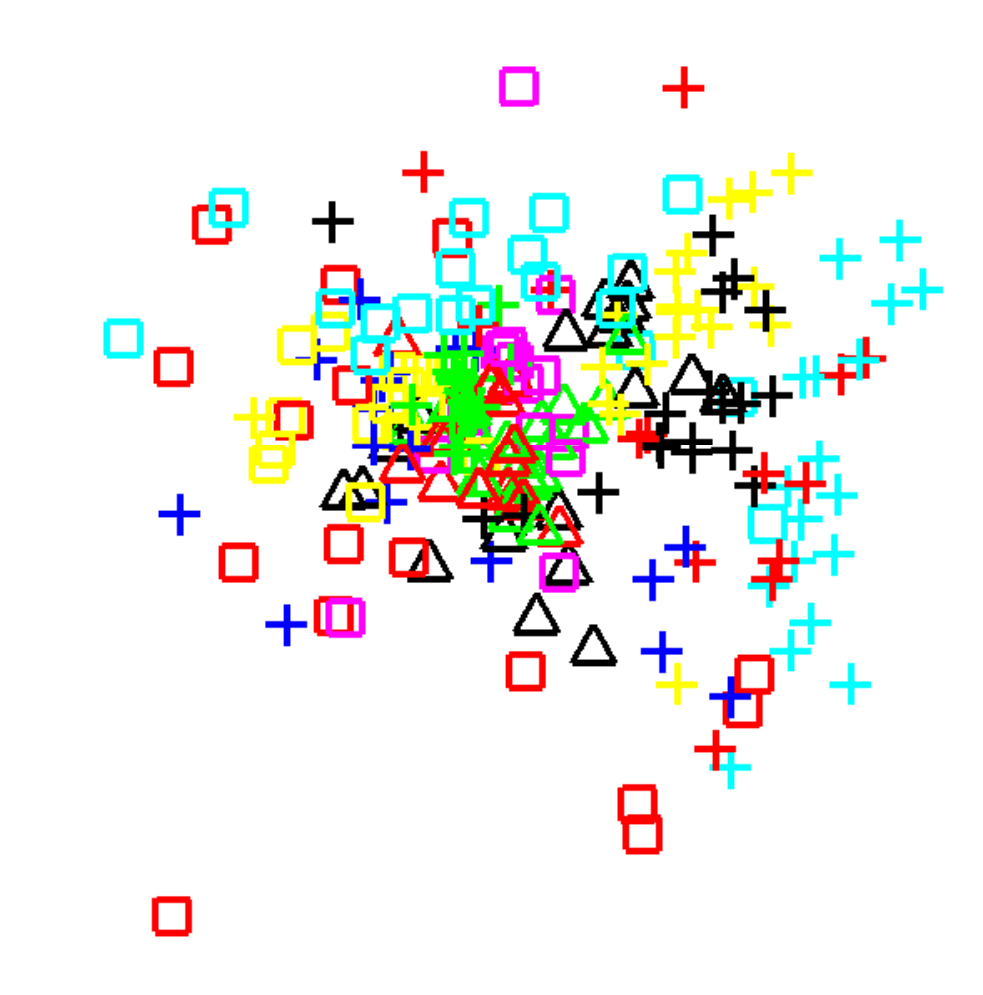} &
		   \includegraphics[width=.24\textwidth]{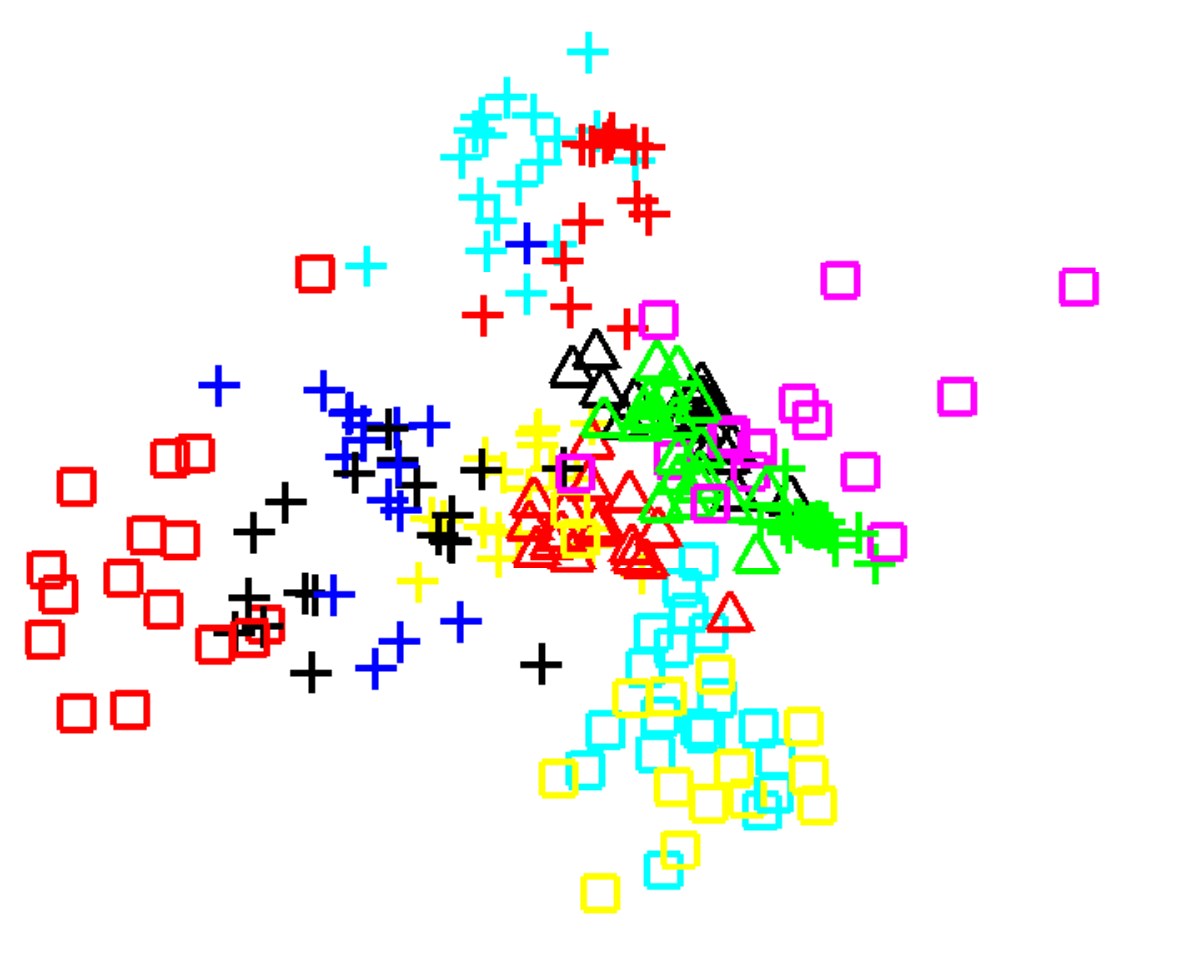}\\
		    \small{(a) $\ltwo$ metric in $\preshapesf$} & \small{(b) $\ltwo$ metric in $\shapesf$, with}\\
		    \small{without registration} & \small {SRNF registration}
	\end{tabular}
	
	\includegraphics[width=.48\textwidth]{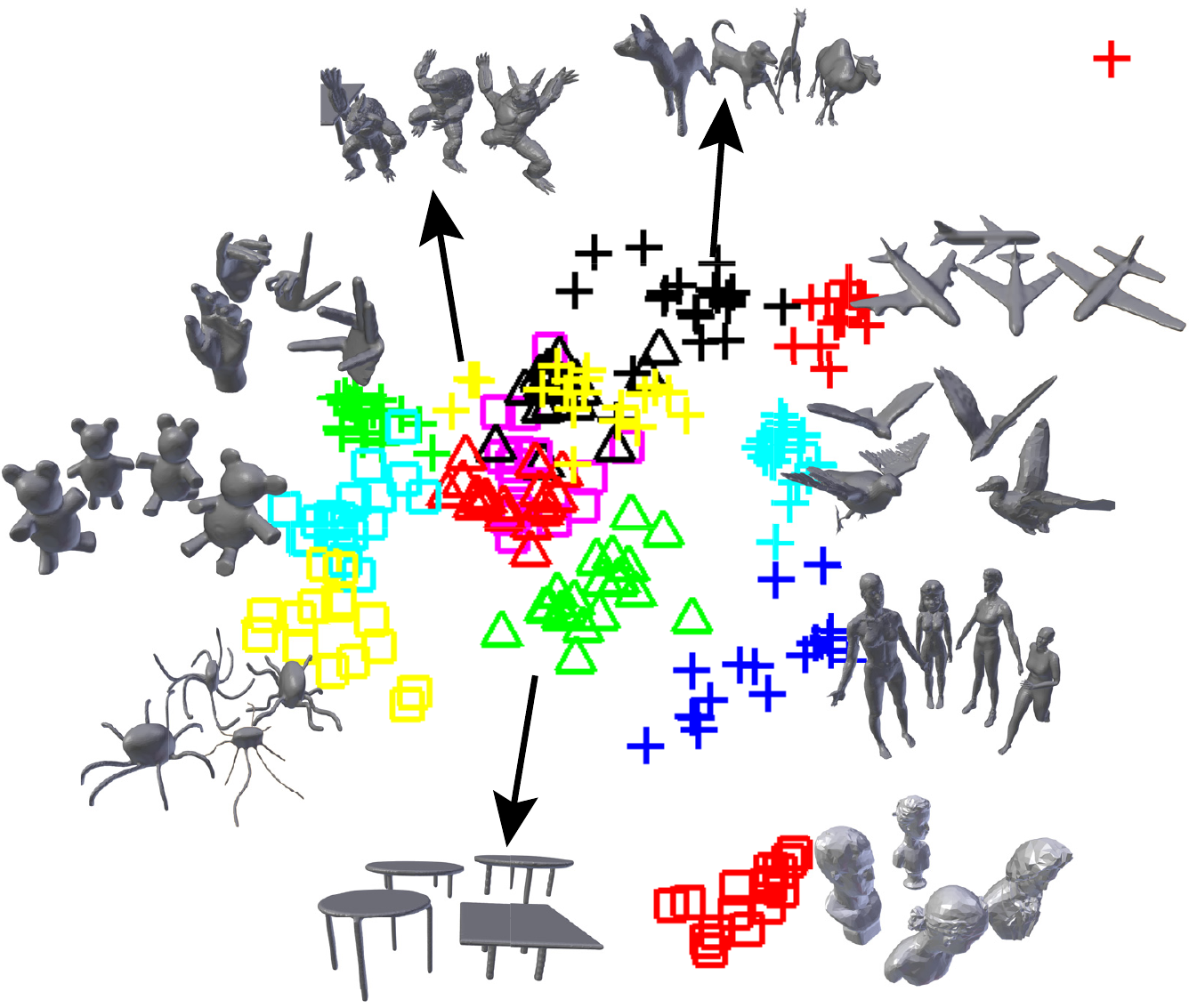} \\
	\small{ (c) $\ltwo$ metric in SRNF shape space, $\shapesq$,  with elastic registration. }
		 
    \caption{\label{fig:shrec_mds} MDS plot of the pairwise distance between every pair of models in the SHREC07 watertight database.  (Image best viewed in color.)}
\end{figure}

The length of a geodesic, or deformation energy, between a pair of surfaces in the shape space is a measure of
dissimilarity that can be used for unsupervised as well as supervised classification of 3D objects. To demonstrate this, we use the SHREC07 watertight 3D model benchmark~\cite{giorgi2007shape}, which is composed of $400$ watertight 3D models evenly divided into $20$ shape classes. We only consider the $13$ classes that are composed of genus-0, triangulated meshes. First, we compute their spherical parameterizations~\cite{laga2004geometry,laga2006spherical,Kurtek:CGF12063}, and then normalize them  for translation and scale. Thus, these surfaces become elements of the preshape space $\preshapesf$. Next, we map the surfaces to the space of SRNFs, $\preshapesq$ and compute their pairwise distance matrix. That is, for a given pair of surfaces $f_1$ and $f_2$ and their respective SRNF maps $q_1$ and $q_2$, their dissimilarity is given as 
$$
	d((f_1, f_2) = d_{\shapes}([f_1], [f_2]) =  \min_{O, \diffeo} \| q_1 - O(q_2 \circ \diffeo) \|^2.
$$
For comparison, we also compute the pairwise distances between the shapes using (1) the $\ltwo$ distance in $\preshapesf$, i.e. without registration, and (2) the $\ltwo$ metric in $\shapesf$ after elastic registration using the SRNF framework. Figure~\ref{fig:shrec_mds} shows the multidimensional scaling (MDS) plot of these three pairwise distance matrices. As one would expect,  when using the $\ltwo$ metric in $\preshapesf$, the 3D models are all spread across the space (Figure~\ref{fig:shrec_mds}-(a)). Clusters start to emerge when using the $\ltwo$ metric in $\shapesf$ with elastic registration performed using the SRNF framework (Figure~\ref{fig:shrec_mds}-(b)). When using the $\ltwo$ metric in $\shapesq$, which is equivalent to partial elastic metric (Figure~\ref{fig:shrec_mds}-(c)), we can clearly see that the 3D models are clustered by shape classes. This suggests that the SRNF framework is suitable for 3D shape classification and clustering. We refer the reader to~\cite{laga2016numerical} for a  detailed  analysis of the performance of  SRNFs in  classification and clustering.

\subsection{Random 3D model synthesis}

Finally, we consider the problem of synthesizing arbitrary 3D models that are similar to, but note exactly the same as, a given collection of 3D models.  We  consider a collection of $398$ human shapes~\cite{hasler2009statistical} composed of multiple subjects in different poses. The first subject has $35$ different poses including a neutral one. All other subjects have a neutral pose and a few other poses.  We first compute the SRNF representations of these surfaces, perform statistical analysis in the SRNF shape space $\shapesq$, using standard linear statistics such  as Principal Component Analysis, and finally map the results to the surface space, $\Space{F}$, using the proposed SRNF inversion algorithm. Formally, if $\bar{q}$ is the mean and $V_i, i=1, \cdots, k$ are the leading eigenvectors in the space of SRNFs, then any arbitrary SRNF map $q$ can be written as:
$$
	q = \bar{q} + \sum_{i=1}^k \lambda_i V_i,
$$
where $\lambda_i$ are real coefficients. An arbitrary shape can then be synthesized by finding a surface $f$ such that its SRNF is $q$. This can be efficiently computed using the SRNF inversion procedure described in Section~\ref{sec:srnf_geodesics} and detailed in~\cite{laga2016numerical}.

\begin{figure}[t]
    \centering

    	\includegraphics[width=.48\textwidth]{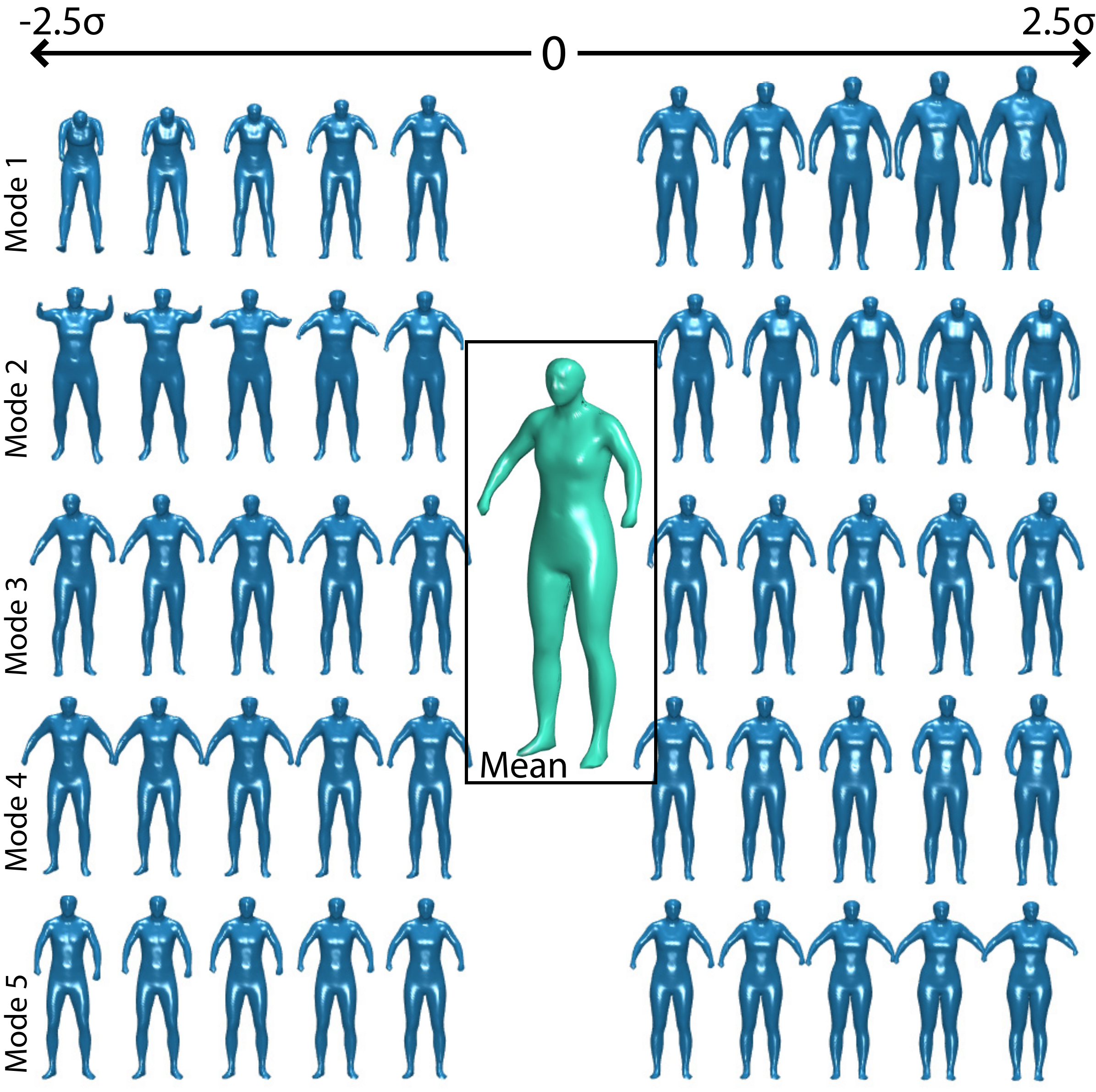} \\
	
    \caption{\label{fig:mean_modes_shape} Mean (central shape) and first five leading modes of variations, computed using SRNF inversion,  of a collection of 3D human body shapes in different poses and belonging to different subjects. }
\end{figure}

\begin{figure}[t]
    \centering

    	\includegraphics[width=.48\textwidth]{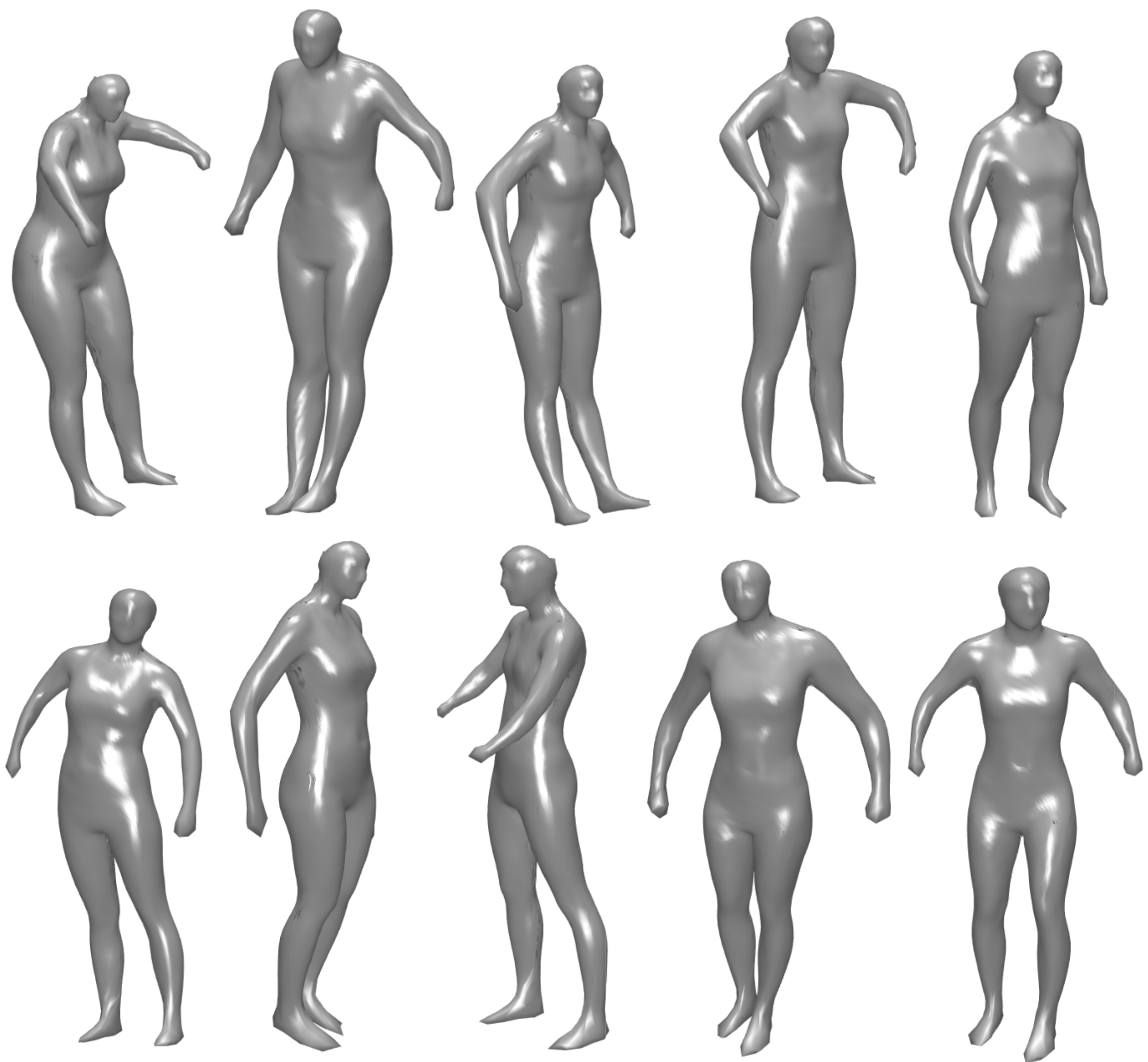} \\
	
    \caption{\label{fig:random_shapes} Ten arbitrary 3D human body shapes automatically synthesized by sampling from a Gaussian distribution fitted to a collection of human body shapes in the SRNF space. }
\end{figure}

Figure~\ref{fig:mean_modes_shape}-(a) shows the mean shape and the first five modes of variation. Figure~\ref{fig:random_shapes}, on the other hand, shows $10$ randomly syntehsized human body shapes. Observe that the statistical model can generate shapes of arbitrary subjects in arbitrary poses.

\subsection{Other applications}
Elastic registration,  geodesics and statistical summaries are building blocs to many applications in computer graphics, computer vision, and medical image analysis. The methods presented in this chapter are able to perform most of  these tasks.   As a consequence, the framework has been used in various applications.  In medical applications,  Kurtek et al.~\cite{kurtek:mi2011,kurtek_mmbia2012} used the Euclidean distance in the space of Q-maps for shape-based analysis of 3D subcortical structures in the brain in order to detect attention-deficit hyperactivity disorder (ADHD).  Joshi et al.~\cite{joshi2016surface}, on the other hand,  used the SRNF representation for shape-based hippocampal modeling in Alzheimer's disease.  
In particular, Joshi et al. showed that the elastic metric preserves important anatomical features when matching hippocampal structures and thus is potentially useful for accurately capturing the biological homology. The approach naturally provides a statistical framework for computing means and covariances in the shape space, which can be used for modeling the biological variability of subcortical structures. They observed a slight overlap of the statistical group difference results, obtained using SRNF, when comparing with the SPHARM approach~\cite{styner2006framework}. Joshi et al.'s results, however, reveal additional effects due to disease or progression in different subregions of the hippocampal anatomy. One possible explanation is that the deformations generated using re-parameterization-invariant elastic metrics also account for the variability due to parameterizations, and thus may contain a richer description about the shape variability of the anatomical structures.  Samir et al.~\cite{samir2014elastic} used the SRNF framework for elastic shape analysis of cylindrical surfaces and applied it to  3D/2D registration in endometrial tissue characterization and also to simulate the deformation of such tissues~\cite{kurtek2016statistical}. 

Elastic shape analysis, in particular SRM-based elastic shape analysis~\cite{kurtek2010novel,kurtek:pami2012}  has been later extended to include landmark constraints in the analysis process~\cite{Kurtek:CGF12063}. The landmarks can be used either to guide  registration in the presence of large and complex elastic deformations~\cite{Kurtek:CGF12063}, or as anatomical features that one needs to preserve when  generating patient-specific atlas of anatomical organs such as carpal bones~\cite{banerjee2015generation}.  


\section{Summary and perspectives}
\label{sec:summary}

We have reviewed  in this chapter the fundamentals as well as the state-of-the-art methods for non-rigid 3D shape analysis. We have seen that there are three  fundamentally inter-related problems, which are  correspondence (registration), geodesic deformation, and statistical summarization of shapes that undergo elastic deformations. Thus, these problems are better solved jointly using the same metric that captures (and quantifies) physical deformations, which are bending and stretching.  We have particularly seen   three types of metrics, mainly (1) the Euclidean distance or $\ltwo$ metric in the space of shapes $\preshapesf$, (2) the pullback of elastic metrics, including the elastic metric in the space of shells, and (3) the $\ltwo$ metric in the space of Square-Root Normal Fields (SRNF). While metrics in the first  class  are easy to use and have lead to some of the major developments in statistical shape analysis including the active shape models  as well as the morphable models, they are limited to 3D objects that do not undergo large bending and stretching. Moreover, they are not reparameterization invariant and thus the correspondence problem should be solved in a pre-processing step using a different approach and a different optimality criteria.

The second class of methods capture the true deformations of shapes and thus can generate natural geodesics and shape summaries. They are, however, computationally very expensive; computing one geodesic may take hours in some cases since they require solving a complex optimization problem. In fact, only a few methods (\eg  M-reps and Lie bodies) have closed-form formulas for computing geodesics and geodesic distances. Other methods, compute them by solving expensive optimization problems.

Square Root representations, on the other hand, are particularly promising since they map 3D shapes onto an Euclidean space  where the $\ltwo$ metric  is equivalent to the partial elastic metric in the preshape and shape spaces $\preshapesf$ and $\shapesf$. Thus geodesics   become straight lines.  One can then perform all the computations in the space of SRNFs using standard vector calculus and only map the results back to the space of surfaces for visualization purposes.  Moreover, diffeomorphisms act by isometries on the elements of the SRNF space under the $\ltwo$ metric. Thus, one can solve jointly the elastic registration problem under the same metric. The main limitations of these methods is that the $\ltwo$ metric on the space of SRNFs is only equivalent to the partial elastic metric with fixed weights for the bending and stretching terms. The existence of such nice representation for arbitrary weights is in fact unknown and can be a promising direction for future research.

Finally, despite the rich literature in this field, there are several fundamental problems and challenges that require further research. Below we summarize some of the open challenges that we believe require further investigation. 

\vspace{6pt}\noindent \textbf{Topological and structural variabilities.  } 
While a significant progress has been made in studying geometric variability (driven by the computer vision and medical imaging communities),    analyzing structural variability of man-made 3D shapes, which is receiving a growing interest by the graphics community,  is still at its infant stage.  Structural variability  is highly related to the semantics and functionality of  shapes~\cite{laga_acmtog2013}. Methods that studied it  (driven by the computer graphics community and  motivated by the need to easily  create 3D models and enrich collections),  do not come with a concept of structure space and a metric for performing true statistics on structure. On the other hand, discrete and  continuous representations (driven by the computer vision, medical and biological imaging communities) lead to well defined shape spaces and metrics.  These methods, however,  perform poorly on 3D shapes that are composed of multiply interacting components, \eg  subcortical structures in the brain.

\vspace{6pt}\noindent\textbf{Multiply-interacting shapes. } Objects are not isolated in nature. Their shape is often influenced by their surroundings and by objects with which they interact. Examples include   anatomical surfaces, \eg  subcortical structures in human brains,  or man-made objects whose shape is influenced by their surroundings and by the actions to which they are subject to. Graphics papers that explored structural variability in man-made 3D shapes capture some of these interactions. Extending these techniques  to natural objects (\eg  subcortical structures) is an interesting direction of research to explore. We believe that developing  a  mathematical representation of the space of structures  and capturing and modeling the way shapes of multiple objects co-vary  are  open challenges that  require  further investigation. Solving them  will open perspectives for new applications in medical image analysis, biology and 3D modeling in computer graphics.

%

\vspace{6pt}\noindent \textbf{Correspondence. }  Despite the large amount of research on this particular topic, correspondence between 3D shapes still remains an open challenging problem, particularly in the presence of large elastic, topological and structural variations. Methods where correspondence and analysis are solved jointly under the same metric provide promising results, but still   fail when analyzing 3D shapes that are semantically or functionally similar but significantly differ in their geometry or structure.  This problem will be  simplified if semantics and functionality of shapes and their parts can be  incorporated into the analysis. Recent works that relate structure to functionality~\cite{laga_acmtog2013,zheng2013smart} provide a step forward in this direction. However, the main challenge when dealing with man-made 3D objects comes from the fact that correspondences are not one-to-one and thus cannot be represented with diffeomorphisms or other nice mappings.  Recent works on graph and tree-graph statistics~\cite{feragen2013toward} may provide part of the solution, although having a joint model that captures both geometric and structural variabilities remains an open problem.

\ifCLASSOPTIONcompsoc
  \section*{Acknowledgments}
\else
  \section*{Acknowledgment}
\fi
The 3D human shape models used in this chapter have been kindly provided by Nils Hasler.  The other models are from the TOSCA~\cite{bronstein2008numerical} and SHREC07 datasets~\cite{giorgi2007shape}. 

\ifCLASSOPTIONcaptionsoff
  \newpage
\fi



\bibliographystyle{IEEEtran}
\bibliography{references}

%

%
%

\end{document}